\documentclass[12pt,preprint]{aastex}
\shorttitle{Massive Binaries R136}
\shortauthors{Massey, Penny, \& Vukovich}
\begin{document}

\title{Orbits of Four Very Massive Binaries in the R136 Cluster}

\author{Philip Massey\altaffilmark{1}}
\affil{Lowell Observatory, 1400 West Mars Hill Road,
Flagstaff, AZ 86001}
\email{massey@lowell.edu}
\author{Laura R. Penny}
\affil{Department of Physics and Astronomy, College of Charleston,
66 George Street, Charleston, SC 29424}
\email{pennyl@cofc.edu}
\and
\author{Julia Vukovich\altaffilmark{2}}
\affil{Lowell Observatory, 1400 West Mars Hill Road,
Flagstaff, AZ 86001}

\altaffiltext{1}{
Based on observations made with the NASA/ESA Hubble Space Telescope, 
obtained 
at the Space
Telescope Science Institute, 
which is operated by the Association of Universities for Research in Astronomy, Inc., under NASA
contract NAS 5-26555. These observations are associated with proposal 8217.}

\altaffiltext{2}{
Participant in the National Science Foundation's Research Experience for Undergraduates (REU) program,
summer 2001.  Current affiliation: Wichita State University}

\begin{abstract}

We present radial velocity and photometry for four early-type,
massive 
double-lined spectroscopic binaries in the R136 cluster.
Three of these systems are eclipsing, allowing orbital inclinations to 
be determined.  One of these systems, R136-38 (O3~V + O6~V), has one of 
the highest masses ever measured, 57$\cal M_\odot$, for the primary.
Comparison of our masses with those
derived from
standard evolutionary tracks shows excellent agreement. 
We also identify five other light variables in the R136
cluster which are worthy of follow-up study.
\end{abstract}


\section{Introduction}

Empirical checks on the mass-luminosity relation and on how well the 
evolutionary models match reality are sorely lacking for high mass 
($>20\cal M_\odot$) stars, even near the zero-age main-sequence (ZAMS).
It has long been recognized that evolutionary tracks (past the ZAMS) are
affected by mass-loss (see, for example, Brunish \& Truran 1982), and it
is now well recognized that in particular the models are highly sensitive
to how mixing of material to and from the core is treated (Maeder \& Conti 1994), with the most recent work emphasizing the importance of rotation
in this regard (Meynet \& Maeder 2000; Maeder \& Meynet 2000; Heger \& Langer 2000; Heger, Langer, \& Woosley 2000).  Yet these
models, untested as they may be, provide the only means for linking a 
variety of observational studies to astrophysical interpretation, such
as the determination of the initial mass function (Massey 1998) and the
determination of turn-off masses in clusters containing massive stars
(Massey, Waterhouse, \& DeGioia-Eastwood 2000,
Massey, DeGioia-Eastwood, \& Waterhouse 2001).

Herrero et al.\ (1992) first called attention to a significant
{\it mass discrepancy} between the masses derived from modern stellar
atmosphere models, and that inferred from stellar evolution codes, for
the highest mass stars, in the sense that the evolutionary tracks predict
a mass as much as 2 times larger. (See also Herrero, Puls, \& Villamariz 2000.)
Burkholder, Massey, \& Morrell
(1997) recently tried to resolve this mass discrepancy by using data
on massive spectroscopic binaries.  They found good agreement between
the binary masses and the evolutionary tracks up to 25$\cal M_\odot$.
Some higher mass systems did show significant lower masses than the
evolutionary tracks would suggest, but these systems were either at or
near their Roche lobes, suggesting that significant mass might have been
lost from the system.  (See also Penny, Gies, \& Bagnuolo 1999.)

Massey \& Hunter (1998) recently obtained spectra of 65 of the bluest,
most luminous stars in the R136 cluster located at the heart of the 30
Doradus nebula in the LMC.  The majority of these stars proved to be of
spectral type O3, the hottest, most luminous, and most massive stars known.
Four of the stars showed widely spaced, double absorption lines, indicative
of spectroscopic binaries caught at favorable phases.  Since the evolutionary
masses of these stars were very high, it was thought that these would be
excellent candidates for additional study, in the hopes that orbit solutions
would resolve the mass discrepancy once and for all.

\section{Observations and Reductions}

\subsection{The Data}

The data were all obtained as part of a Cycle~8 {\it HST} program,
GO-8217. We utilized 30 orbits, organized into 15 ``visits" of
2 consecutive orbits, with each visit separated by
carefully planned intervals.  We used the Space Telescope
Imaging Spectrograph (STIS) in imaging mode for photometry of our
binaries, and with a medium-dispersion grating for spectroscopy of each
binary.

Each visit
began with a pair of short images (total integration time of 2.2 seconds)
obtained with the ``long-pass" filter (which cuts off all light $<5500$\AA)
and centered on the middle of the cluster. The 28 arcsec by 51 arcsec
field of view always contained our four binaries, but changes in the
spacecraft roll angle resulted in some of the out-lying R136 stars
either being included or not.  The spatial scale was 0.05 arcsecs per
pixel.

After centering up on a fairly isolated
offset star (Melnick 34 = R136-8; see Fig.~1 of Massey \& Hunter 1998),
the telescope was precisely offset to each of the four binaries in
turn, in order to obtain spectra with the G430M grating centered at
$\lambda 4451$.  These exposures covered the
wavelength range 4310\AA\ to 4590\AA\ at a dispersion of 0.28\AA\ per pixel,
with a nominal 1.5-pixel resolution of 0.42\AA\ (28 km s$^{-1}$).  This
wavelength region was selected to include as many spectral lines
as possible in a single wavelength setting; i.e.,
H$\gamma$, He~I~$\lambda 4471$, and
He~II~$\lambda 4542$. We used
the 0.2 arcsec by 52 arcsec slit.  Wavelength calibration was obtained
for each exposure using the default {\it HST}/STIS scheme, resulting in
a new calibration exposure prior to the first spectroscopic exposure for
each orbit.

Near the end of each visit, another pair of short-exposure images 
(total integration time 2.6~seconds) were obtained to continue
the photometric monitoring.  The total elapsed time for each
visit was 3.2 hours.

In order to maximize our phase coverage for all periods of interest, 
we adopted
a clever scheme developed by Abi Saha for observing Cepheids as part of
the distance-scale key project: we designed our program so that each
visit was separated in time according to a geometrical progression,
corresponding to 

$$ {\rm gap}(n){\rm [days]} = 0.5 * 1.175^n, n=1, 14 $$

The multiplicative and geometrical factors are chosen to assure good phase
coverage for the shortest and longest periods of interest, given a
predetermined
number of observations.
Simulations showed that this would
provide uniformly good phase coverage for periods from $<1$ day through 30
days.  The longest interval between successive visits was 3.7 days, and
the shortest interval was 0.5 days.  The observations spanned an observing
``season" of 24.4 days.  In arranging the order of our visits, we put all
the shorter intervals near the middle of the 24 day sequence, with progressively
longer intervals to either end.  The result of this is that any eclipses would
tend to be found in the middle of the sequence, as we will find in
Sec.~\ref{sec-otherlcs} is indeed
the case.

\subsection{Photometry}

For the STIS images, we used the ``standard pipeline" flat-fielded images prior to cosmic-ray (CR) rejection. They
had been observed in ``CR-split" mode, resulting in
a pair of images, each of 1.1--1.3~sec duration.  
We chose to do photometry of all 60 images,
and rejected CR events by comparing the photometry from each of
the CR-split images.  If the photometry agreed to within the photometric
errors (3$\sigma$), 
we averaged the results; if not, we assumed that the smaller magnitude
was spurious due to a CR event.  The photometry was measured using IRAF's
aperture photometry routine, with a radius of 2 pixels and sky being 
determined from the modal value within an annulus lying 5-10 pixels from
the star's center. The point-spread-function has a full-width at half-radius
of 1.7 pixels, and we found that the centers were well determined
by a Gaussian fit to the
radial profile.

In order to provide more accurate differential photometry for our binaries,
and to search for other light variables, we performed photometry of 59 of
the brightest, most isolated stars in the cluster, including our four
binaries.  Of the 59 stars, one happened to lie outside the field-of-view
for two of the visits, due to the slightly changing roll-angle of the
spacecraft over the course of the observations.
After eliminating possible variables (based upon whether agreement
frame-to-frame was within the photometric errors), we found that there were
indeed photometric zeropoint changes of $\sim 0.04$ mag from frame-to-frame,
with the drifts being secular rather than random in time: i.e., over
the course of several days the zero-point would drift upwards and then
downwards again. This is consistent with the claim
 that ``instrumental
[in]stability" limits absolute photometry with STIS
to 5\% (Table 16.3 in Leitherer et al.\ 2001). In any event, our use of 
STIS as
an N-star photometer (with N$\sim58-59$) resulted in photometry with high
precision for temporal changes.

We have applied a single adjustment to the photometric zero-point to
have the photometry roughly match $V$; a single value worked well as
the colors of these stars are all very similar (Hunter et al.\ 1997).

\subsection{Spectroscopy}

Owing to the extreme crowding in the R136 cluster, the spectra for
each of the four binaries had to be re-extracted with a small
aperture centered on each target, with care being taken
to select ``clean" sky on either side of the slit.  We actually reduced
our spectra two ways.  For one version, we began with the pipeline final
two-dimensional spectrum (flat-fielded, CR-removed, geometrically-corrected,
and wavelength calibrated) and simply re-extracted the spectra of our
binaries.  For the other, we began with the flat-fielded version, and
extracted our spectra using the usual IRAF routines; for these we also
applied our own dispersion solution determined from the wavelength-calibration
frames.  We found in practice that our own reductions provided better
signal-to-noise.

In measuring the spectra, we fit double lines using two Vogt functions;
this provided excellent fits to even blended double lines, and avoids
issues with pair-blendings, always a concern with orbit solutions of
broad-lined stars (see discussion in Burkholder et al.\ 1997).
The single-lined phases were measured with a single Vogt function for
consistency.  Both versions of the spectra were measured independently,
and the results compared; no systematic differences were found, and the
comparison proved useful mainly for eliminating any spurious measurements
due to noise spikes. Our measuring error was 5-10 km s$^{-1}$.

\section{Results}

We provide the photometric data and radial velocity data in Tables~1 and 2.
In this section we describe what we learned from the spectra of each 
binary, and provide details of the orbit solutions and analysis of
the light-curve information.  We also report the discovery of five more
light-variables in the R136 cluster. We begin by explaining our methods.

\subsection{Methodology}

For the period searches, we used a version of the Lafler \& Kinman (1965) routine, which relies upon point-to-point smoothness
after the data (either
photometric or radial velocity) are sorted in phase according to a
trial period.  For the orbit solutions, we initially solved
for each component independently using a modified version of the
differential corrections program of Wolfe, Horak, \& Storer (1967),
fixing only the period.
In all cases we found solutions consistent with
circular orbits, which we expect given the
short periods and high masses (see Sect.~\ref{sec-disc}). 
We then determined the best values of 
the orbital
semi-amplitudes $K$ ``center-of-mass" $\gamma$-velocities by running a non-linear
least-square routine based upon the grid-search program of
Bevington's (1969), with the eccentricity, time of conjunction, and
periods fixed
for the two components.  This then maximizes the precision of our
determination of $\gamma$ and
$K$.  As a reminder, we expect that the $\gamma$-velocities of the two
components of a massive binary will differ due to the photospheric
outflow velocities of the stellar winds (Massey \& Conti 1977).
We characterize the agreement between the orbit and velocity data by
R1, computed from the goodness-of-fit
$\chi$.

The spectral types and magnitude difference between the components can
be determined from the best double-lined phases.  Our spectral types
differ slightly from those of Massey \& Hunter (1998), as these STIS
spectra have considerably higher resolution and better signal-to-noise.
The relative strengths of He~II~$\lambda 4542$ and He~I~$\lambda 4471$
helped establish the spectral class; the absolute visual magnitudes 
were consistent with
all of these stars being dwarfs.  The magnitude difference could be
measured from the relative fluxes of the H$\gamma$ lines at double-lined
phases, as the equivalent width of H$\gamma$ is fairly insensitive to
$T_{\rm eff}$ for dwarfs this hot.

We measured the projected rotation velocity $v \sin i$ for each component
by using the appropriate model atmosphere lines and convolving these
with rotational velocities until we obtained the best fit to a line.
The model atmosphere code is that described by Kudritzki \& Puls (2000),
and we used models computed by PM for stars of similar
spectral types in the LMC.
Given the intrinsic line widths, we found we could not measure rotational
velocities smaller than 90-100 km s$^{-1}$.  In the following, we will
compare the rotational velocities to that expected on the basis of
synchronous rotation, computed using the stellar radii and orbital period.

We thus know a great deal about the physical parameters of these stars,
which is helpful in reducing the number of free parameters in interpreting
the light-curves.  We adopt a distance modulus $(m-M)_o=18.5$, in accord
with Westerlund (1997) and van den Bergh (2000).  Accurate values of
the reddening were determined by Massey \& Hunter (1998) based upon
the mulit-band {\it HST} photometry of Hunter et al.\ (1997).  Thus the
absolute visual magnitudes are known (Table~1 of Massey \& Hunter 1998)
for the combined systems, and, combined with the magnitude differences
measured from the spectra, tells us $M_V$ for each component in the
binary.  The spectral type allows us to assign effective temperatures;
we adopt the scale of Chlebowski \& Garmany (1991). (We will discuss this
in more detail in Sec.~\ref{sec-disc}.)  The effective temperature then
determines the bolometric correction (Vacca, Garmany, \& Shull 1996),
and hence we know the bolometric luminosity of each star.

We used the light curve synthesis code GENSYN (Mochnacki \& Doughty 1972)
to produce model light curves.  Our approach was to make a constrained
fit using as much data as possible from the spectroscopic results as
described above.  The orbital parameters were taken from the spectroscopic
solution, and the physical parameters were estimated from the spectral
classifications of the stars.  We set the stellar temperatures according
to the spectral classification.  We then estimated the physical fluxes
and limb darkening coefficients from tables in Kurucz (1979) flux models
and Claret (2000), respectively.  The observed flux ratios together with
the adopted effective temperatures yield estimates of the ratio of stellar
radii.  Then, for a given input value of the polar radius of the primary
$R_p$, we calculated the secondary radius.  Each trial run of GENSYN was
set by two independent parameters, the system inclination $i$ and primary
polar radius. For each run, we attempt to match two observables: the
absolute visual magnitude of the system and the eclipse depths and widths.
The best fit solutions are those with the calculated $M_V$ of the system
that also matched the eclipse depths.  Models that fit the eclipse depths
could be made for a range of inclinations, however the $M_V$ for such
models would greatly diverge from that calculated using the well known
distance and reddening to the LMC.  The quoted errors in inclination
derive from an estimated error of $\pm$ 0.15~mag on our $M_V$ values, which
is based upon the uncertainty 
in the LMC distance modulus (van den Bergh 2000)
and a modest 
error in the photometry and reddening (Massey \& Hunter 1998).

For all four systems, the stars are well contained within their Roche
surfaces.  In the case of R136-39, no eclipses are seen.  However,
the crucial phases where we might expect eclipses lack observations.
Therefore, we only quote an upper limit on the inclination.  Any
inclination above this value would result in eclipses that would be both
too deep and wide to agree with the current observations.

As a further check on the models, we independently estimated the orbital
inclination simply from geometry, after measuring the eclipse depths.
For this, we adopt a modest correction for limb darkening using a linear
coefficient (Al-Naimiy 1978; van Hamme 1993), but ignore reflection and
other second-order effects.  Given the poor sampling in phase-space we
expected only modest agreement with the models, but in fact the agreement
was excellent, giving us high confidence that the orbital inclinations
are well determined.

We estimate the errors on the physical parameters, including
a 10\% uncertainty in the effective temperature scale for
O-type stars (Conti 1988).  Nevertheless, our parameters are well
determined, in large part because the inverse dependence of the stellar
radii on effective temperature is partially canceled by the dependence of
the bolometric correction on effective temperature, and in fact the
uncertainties in $\Delta m$ dominate the errors on the stellar radii.
$$L/L_\odot = (R/R_\odot)^2 \times (T_{\rm eff}/T_\odot)^4$$
\begin{equation}
R/R_\odot =   (L/L_\odot)^{0.5} \times  (T_\odot/T_{\rm eff})^2 
\end{equation}
For solar-type stars, a small error in estimating the effective temperature
$T_{\rm eff}$ would result in a large error in $R/R_\odot$, since there is such
a steep inverse relationship
with $T_{\rm eff}$.  However, for hot stars we must make a very
large correction for the bolometric correction (BC) 
in determining $L/L_\odot$ starting
with $M_V$, and that the BC has a very steep dependence on $T_{\rm eff}$ as
well: 
$${\rm BC} = 27.66 - 6.84 \times \log T_{\rm eff},$$
from Vacca et al.~(1996).  Thus substituting this into equation (1), and
adopting $T_\odot=5770 ^\circ$K, we find
$$R/R_\odot = 871.67 \times 10^{-M_V/5} \times T_{\rm eff}^{-0.632}.$$
By propagation of errors we find then that 
$$\sigma_{R/R_\odot}^2 = \sigma_{T_{\rm eff}}^2 \times (-550.90 \times 10^{-M_V/5}
\times T_{\rm eff}^{-1.632})^2 
\\ + \sigma_{M_V}^2 \times (-401.42 \times 10^{-M_V/5} \times T_{\rm eff}^{-0.632})^2$$

\subsection{R136-38}

Six of the 15 spectra showed double lines at H$\gamma$ and He~II~$\lambda 4542$.
There is a large magnitude difference between the two components, which we
measure as $\Delta$m=1.0$\pm0.2$~mag.  The He~I$\lambda 4471$ line clearly
follows the motion of the secondary, but never showed any component due
to the primary.  Thus we have a good measurement of the motion of the secondary
for both single-lined and double-lined phases.  Occasionally the He~I line
was too noisy to measure. We find that the single-lined
phases follow the motion of the primary very well, but we will not give those
any weight in the orbit solutions.  The primary is of spectral type O3~V,
while the secondary is of spectral type O6~V, as judged by the relative
strengths of He~I and He~II during the double-lined phases.
The photometry shows well pronounced dips of $\sim0.2$~mag indicative of
eclipses.

\begin{figure}

\epsscale{0.70}
\plotone{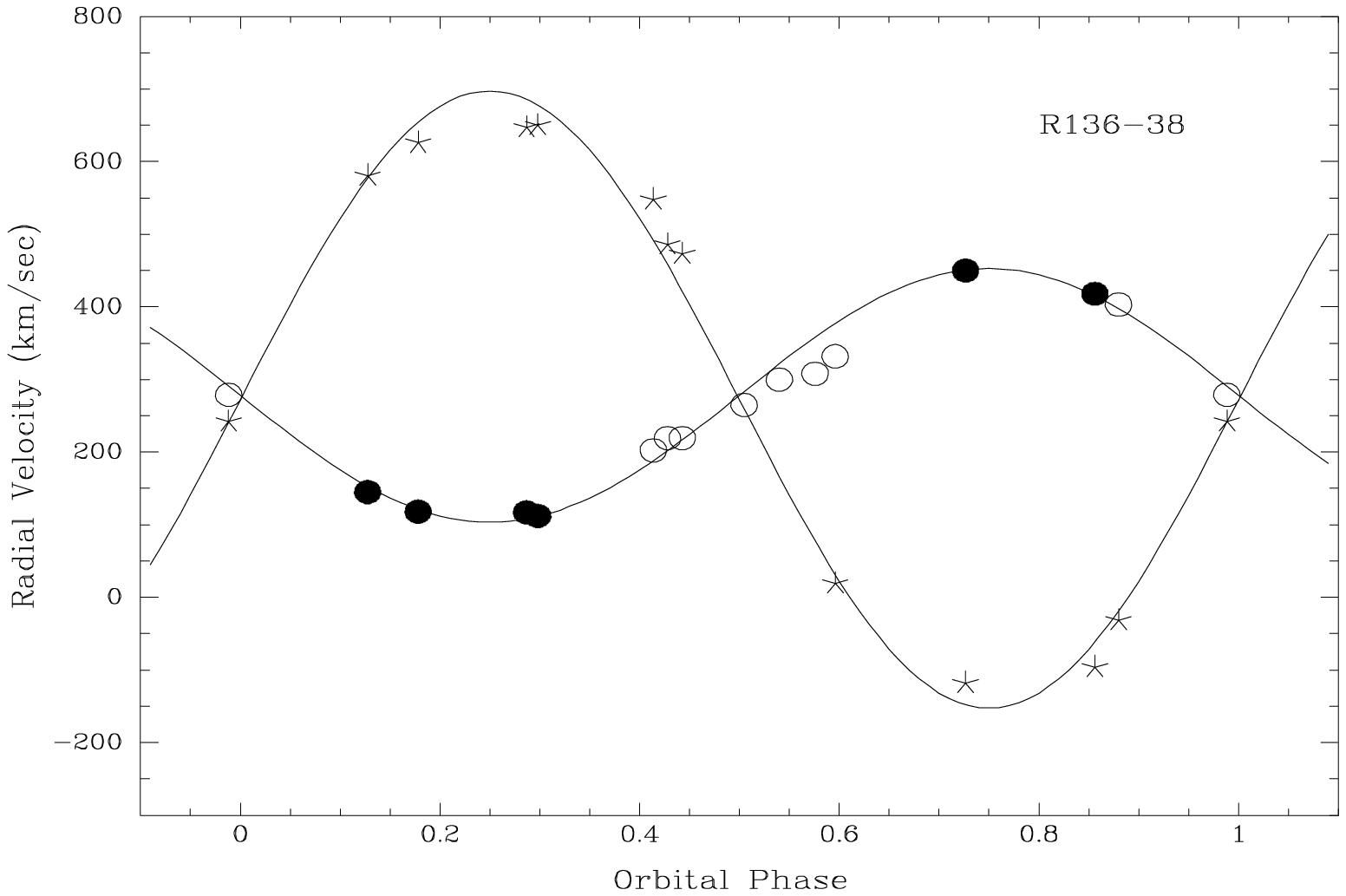}
\plotone{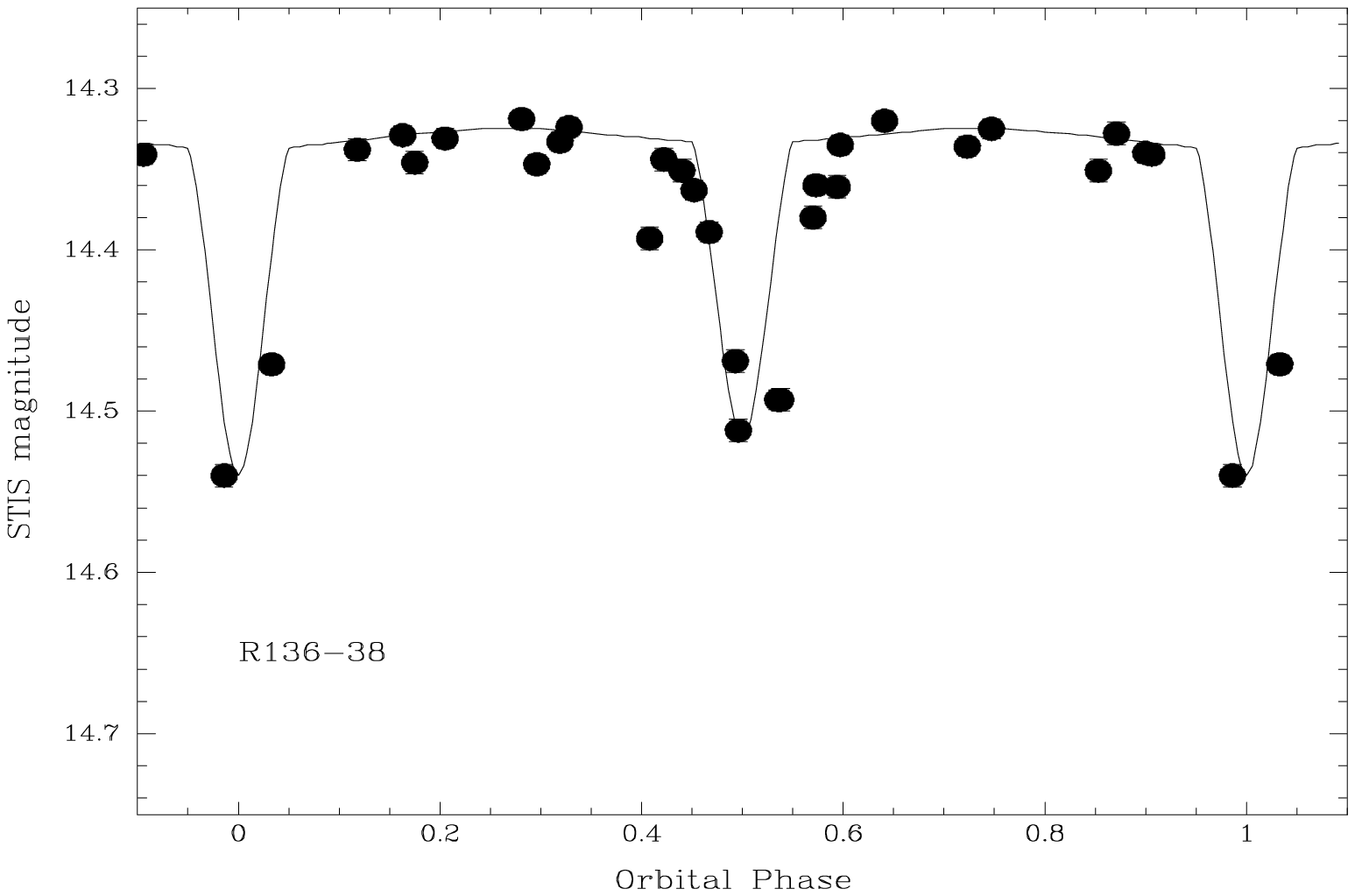}

\caption{Radial velocity and light curves for R136-38.  In the radial
velocity curve, the solid points designate the primary; the asterisks
denote the secondary, and the open circles denote the measurements
at single-lined phases, which were not included in the orbit fit.
In the plot of the light-curve, 
we include the the 1$\sigma$ error bars in the photometry, although for
R136-38 these are comparable to the size of the points.
The solid lines shows the results of our best model fit.
\label{fig:r13638}}
\end{figure}

Period searches on the radial velocity data (both
primary and secondary) and on the photometry all revealed the same
period, 3.39 days. We present the orbital parameters in Table~3, and
show the radial velocity curves and orbit fit in 
Fig~\ref{fig:r13638}.  We find that the rotational velocities are consistent
with synchronous rotation.
Analysis of the light-curve finds a well-determined orbital inclination of
$i=79^\circ \pm1^\circ$.  We show the light curve data, and the model fit,
in Fig~\ref{fig:r13638} as well.

This is a very interesting system, containing stars of extremely high masses,
with the O3~V primary having a mass of 57$\cal M_\odot$.  This is 
higher than the mass determined from any other binary,  
exceeding the mass of
even that of Plaskett's star 
(51$\cal M_\odot$, Bagnuolo, Gies, \& Wiggs 1992).\footnote{There are two other contenders for the recorder-holder.
(1) HD~92740 is a WN7 Wolf-Rayet star in which the absorption lines and
emission lines follow the same orbital motion.  By combining spectra from
six He~I lines, Schweickhardt et al.\ (1999) report the detection of a secondary
absorption spectrum, which they describe as O9~III.  The double-lined
orbit solution yields a mass of $55.3\pm7.3 \cal M_\odot$ for the Wolf-Rayet
star.  (2) HD~93205 is an O3~V + O8~V pair recently studied by
Antokhina et al.\ (2000).  The mass of the O3~V primary was found to be
32-154$\cal M_\odot$, with a most probable value of
45 $\cal M_\odot$. We are indebted to Doug Gies for reminding us of these
systems.}  We compare these to the masses derived from the
evolutionary tracks in Table~3, and find excellent agreement.

\subsection{R136-39}

This system consists of an O3~V + O5.5~V pair.  Of the 15 spectra, 11 show
double lines at H$\gamma$ and He~II $\lambda 4542$, and it is clear the the
He~I $\lambda 4471$ comes purely from the secondary.  The magnitude difference
between the two components is $0.45\pm0.1$~mag.  The period is 4.06~days,
as determined from the radial-velocities.  Inspection of the unphased
photometry did not show any obvious signs of an eclipse, but when phased
according to the radial velocity information we find that there is poor
coverage near the important phases.  This allows us only to assign an upper
limit on $i$, and
our modeling suggests that the orbital
inclination must be less than $75^\circ$ to 
account for the lack of deeper eclipses. 
This places only lower limits on the masses on the
system, of $27 \cal M_\odot$ and $21 \cal M_\odot$.  The line widths are
consistent with synchronous rotation.

\subsection{R136-42}

This system consists of an O3~V + O3~V pair.  Of the 15 spectra, 11 showed
double lines at H$\gamma$ and He~II $\lambda 4542$, and there was no
trace of He~I in any of our spectra.  The magnitude difference between
the two components is modest, $\Delta$m=$0.2\pm0.1$. The light-curve shows
a very deep eclipse (0.5~mag), suggesting that we are viewing this system
at a very favorable inclination.

A period search of both the radial velocity and photometry data yielded
the same value, $P=2.89$ days.  We give the orbital parameters in Table~5,
and show the orbit and light-curve in Fig.~\ref{fig:r13642}.   We find
$i=85.4^\circ$.  The masses are among the highest seen in a spectroscopic
binary, $40\cal M_\odot$ and $33\cal M_\odot$.
The line widths are consistent with synchronous rotation.

\subsection{R136-77}

The last system we discuss consists of two O5.5~V + O5.5~V
stars with equal brightness. Of the 15 spectra, 12 showed double
lines at H$\gamma$, He~I $\lambda 4471$, and He~II $\lambda 4542$.
However, since we could not distinguish the primary from the secondary,
it was necessary to rely upon the velocity {\it differences} between
the two components in order to find the period; a period search of the
photometry yielded identical results, and strong (0.4~mag) eclipses
are evident.  The period was then used to phase the radial
velocity data, and assign velocities to one star or the other.
We designate the slightly more massive star the primary.
The radial velocity data, orbit solution, and light-curve is shown
in Fig.~\ref{fig:r13677}. The line widths are again found to be
consistent with synchronous rotation.

\subsection{Other Light Variables}

\label{sec-otherlcs}

In doing the photometry, we found five additional stars with 
significant photometric variations.  We show the light-curves in 
Fig.~\ref{fig:otherlcs}.  R136-07 (Melnick 39), R136-15 (Melnick 30),
R136-24 (R136a7), and R136-25 all show signatures of what might well be
eclipses.  The star R136-08 (Melnick 34) shows a much more puzzling
behavior, changing by several tenths of a magnitude over the course
of 3 weeks.  The variations appear to be periodic.  The spectrum of 
R136-08 mimics that of a Wolf-Rayet star, although Massey \& Hunter (1998)
argue that it is simply a ``super Of" star whose very high luminosity and
stellar winds result in a spectrum dominated by emission.  Spectroscopic and
photometric monitoring of all five of these stars has been proposed for
Cycle~11 with {\it HST}.

\section{Discussion}
\label{sec-disc}

Our analysis of the four R136 binaries have revealed orbital masses that
are among the highest ever directly measured via this simple application
of Kepler's second law.  We can use these new data to compare to the
masses derived from the evolutionary 
tracks\footnote{We have used
the older Schaerer et al.\ (1993) evolutionary tracks as these
were the
last set made public by the Geneva group that includes normal (rather
than enhanced) mass-loss rates.  Newer models including rotation are becoming
available, but as yet none with the metallicity appropriate to the LMC.
However, we expect that this effect will be small {\it near the ZAMS}, as 
suggested by
Fig.~6 in Maeder \& Meynet (2001).}.  We have included these values in
Tables 3-6, and present the H-R diagrams (HRDs) in Fig~\ref{fig:r136allhrd}.
The agreement between
the masses derived from these tracks, and the actual measured masses,
is excellent for the three cases with eclipses.

In placing the stars in the H-R diagram, we have chosen to adopt the
effective temperature scale of Chlebowski \& Garmany (1991).  This is
similar to the scale given by Conti (1988), who provides a
critical discussion, and concludes that the absolute (but not relative)
uncertainties in the scale are about 10\%.  Since that time, more
modern atmospheric models have been used to analyze a number of O-type
stars in the Milky Way, LMC, and SMC; see, for example, Puls et al.\ (1996).
Such studies led Vacca et al.~(1996) to propose a new
effective temperature scale, which is $\sim$6\% hotter than the
Conti (1988) calibration, and $\sim$3\% hotter than the Chlebowski \& Garmany
(1991) scale for the spectral types discussed in the current paper.  
We note that there are no spectral type to effective temperature scales
determined for stars in the LMC and SMC, and that Vacca et al.\ (1996)
restricted their study to Galactic stars.  Thus refinements in the
effective temperature scale will change the location of the stars in the
HRD but probably the error bars in Fig.~\ref{fig:r136allhrd} are realistic.
Note that we have only included the uncertainty in $M_V$ in estimating
the errors on the evolutionary track masses in Tables 3-6; were we to include
the uncertainty in the effective temperature scale as well, the percentage 
error
would roughly double.

The stars do fall slightly to the left of where we expect in the HRDs. 
Massey \& Hunter (1998) found ages of 1-2~Myr for the R136 cluster, with
the larger value corresponding to the cooler effective temperature scale,
which we have used here.  Yet the components in all four of our binaries
lie on or near the ZAMS, to higher effective temperatures than the
2~Myr isochrones shown in Fig.~\ref{fig:r136allhrd}. 
We do not have a ready explanation for this.
We note that all four of these systems are relatively close pairs.
Comparison of the orbital separations $a \sin i$ with the stellar radii
(both quantities appear in Tables 3-6) reveals that the components are
typically separated by $\sim 2 \times$ the sum of the radii.  This is
sufficiently close that tidal forces have played a significant
role.  We see ready evidence of this in that there must have been some
dynamical evolution for the orbits to be circular and the stars to
be locked in synchronous rotation.\footnote{The time expected for tidal forces to circularize an orbit can be
estimated using equation (2) of Shu \& Lubow (1981);
adopting parameters appropriate to high-mass stars with convective cores
(Zahn 1975, 1977) leads to $\sim 500,000$ years for these systems.}
Such tidal forces may have affected the evolution
of the stars (providing additional heating on the envelopes), and in that
case, these systems are not telling us as much as we
would like about single stars.  It is hoped that some
of the stars identified in Sect.~\ref{sec-otherlcs} may provide example of
massive binaries with longer periods.

Such systems would also help determine if the
new generation of rotating stellar models do better than the standard
non-rotating models.  Maeder \& Meynet (2000) have recently invoked rotation
to explain the discrepancy in masses between the evolutionary tracks and
stellar atmosphere calculations; they note that 
tracks which fail to include rotation may overestimate the mass by as much
as 50\% in cases of high rotation and luminosity.
The short periods of our binaries have resulted in slow rotation due to
tidal forces, leading to little difference
in the masses predicted by non-rotating and
rotating models.

\acknowledgments
We are grateful to Abi Saha for extensive discussions of how best to observe variable objects in order to obtain good phase coverage.  In order to measure
the rotational velocities of the stars, we used the model atmosphere code
of Rolf Kudritzki and his collaborators. We thank 
Deidre Hunter for a critical reading of the manuscript, and Andy
Odell for an interesting conversation about one of these stars.
Kim Venn provided useful suggestions for improving the paper.
Support for REU student JV was provided by NSF grant 9988007.
{\it HST} 
proposal GO-8217 was provided by NASA through a grant from the
Space Telescope Science Institute, which is operated by the Association
of Universities for Research in Astronomy, Inc., under NASA contract
NAS 5-26555.

\clearpage

\clearpage

\clearpage
\begin{figure}
\epsscale{0.70}
\plotone{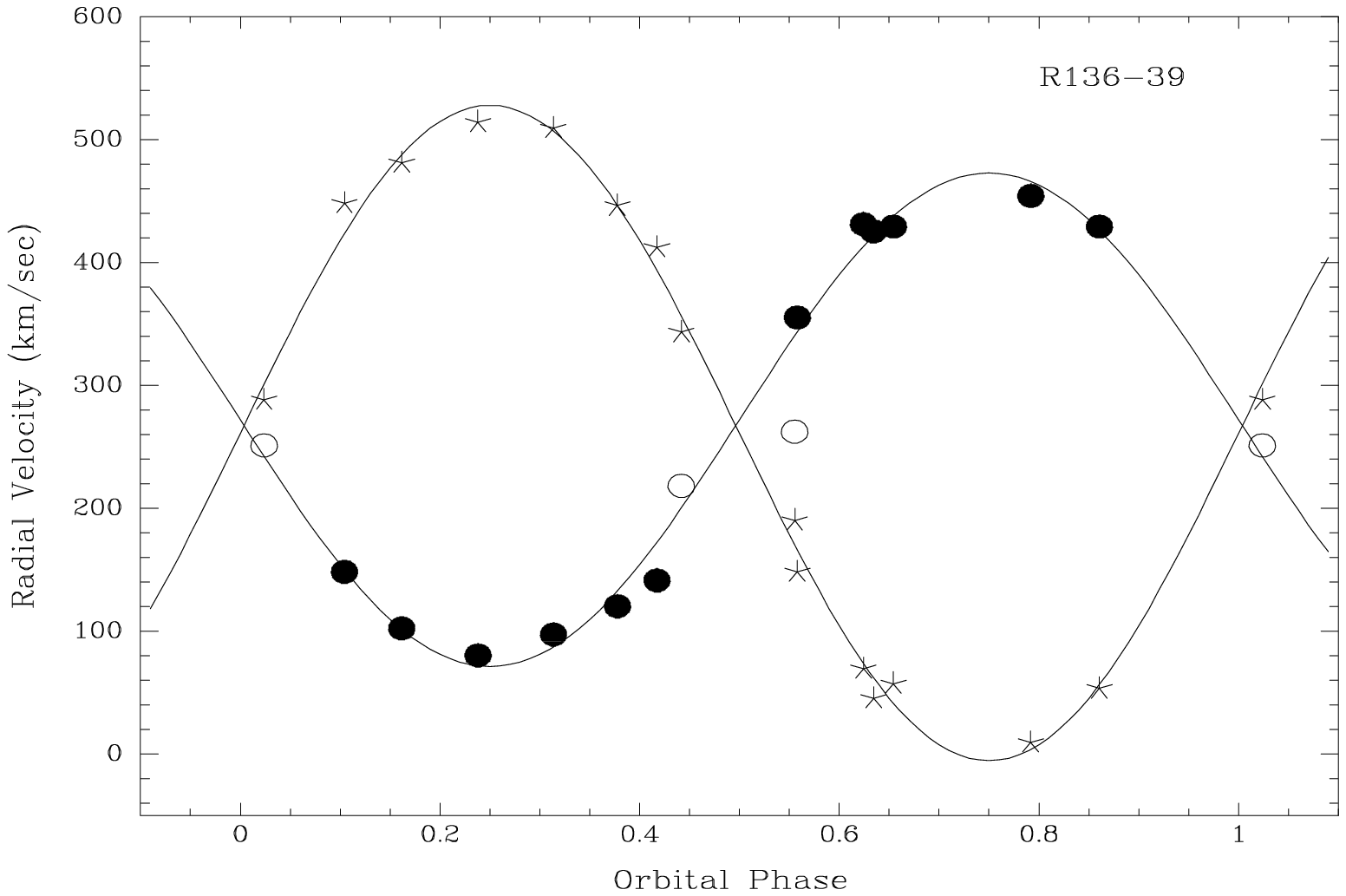}
\plotone{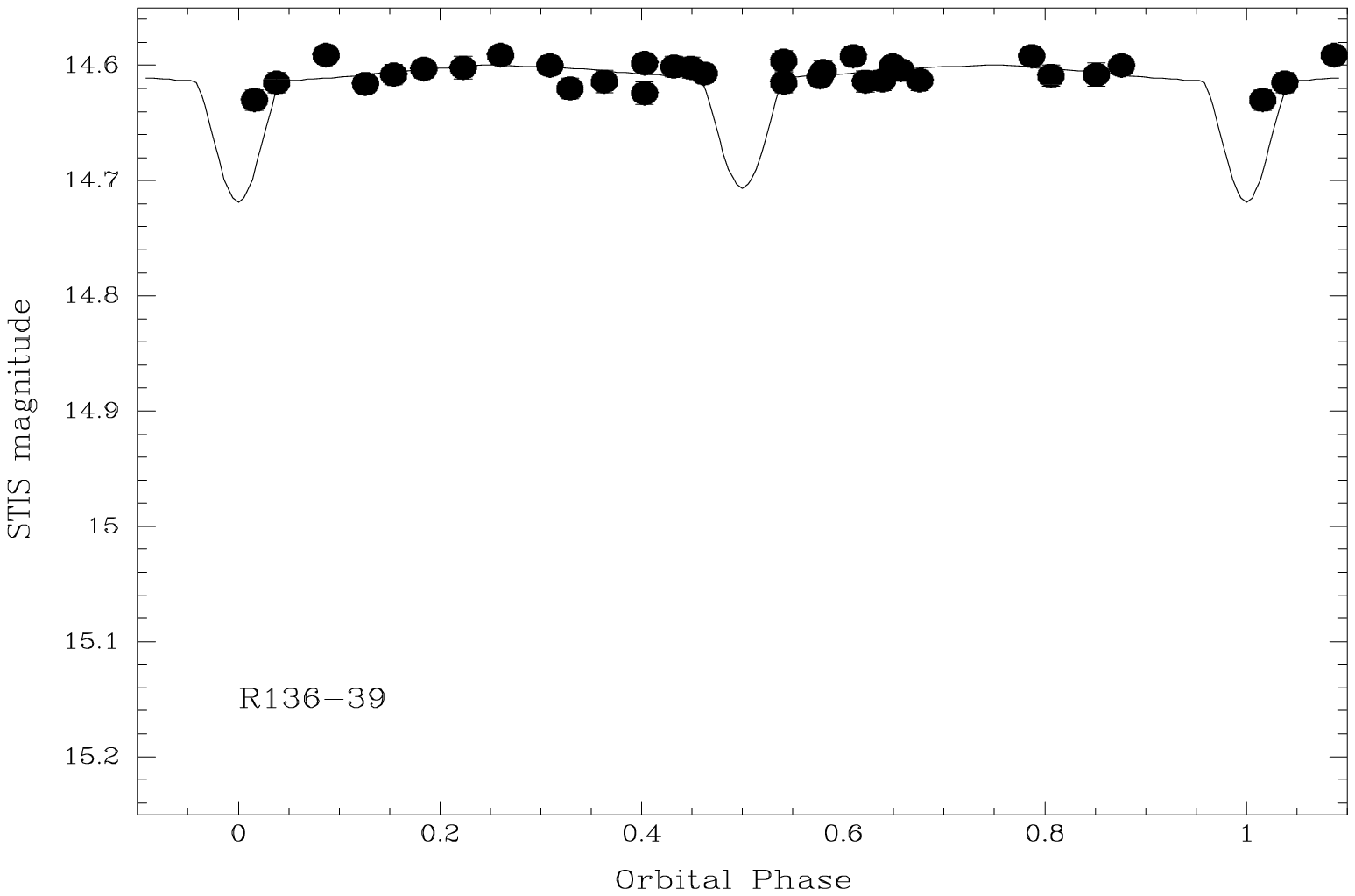}

\caption{Radial velocity and light curves for R136-39; symbols are the
same as in Fig.~\ref{fig:r13638}. The model light-curve fit is for the
maximum possible eclipse depth consistent with the data; we use this
for setting an upper limit on the orbital inclination. \label{fig:r13639}}
\end{figure}

\clearpage
\begin{figure}
\epsscale{0.7}
\plotone{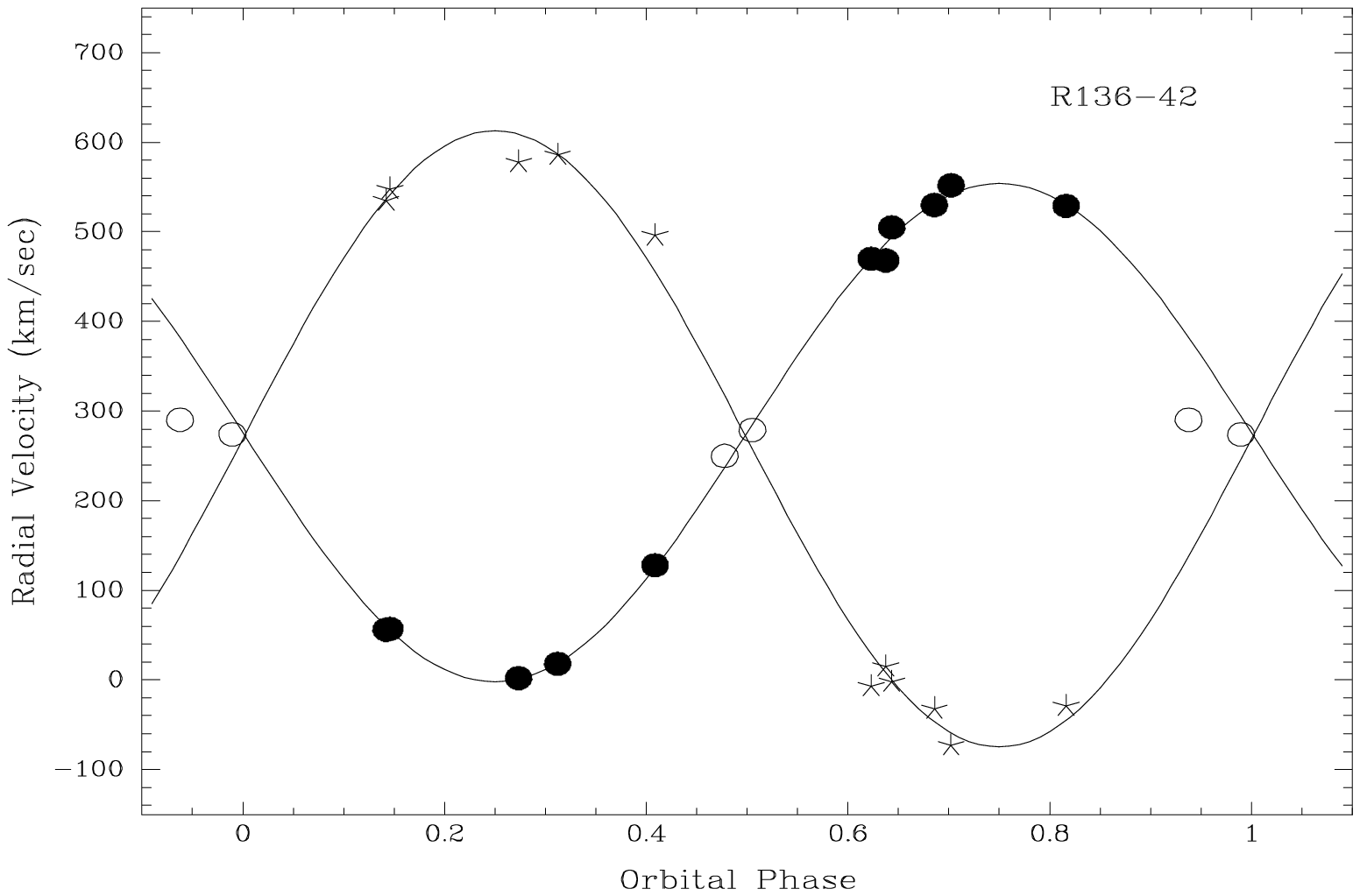}
\plotone{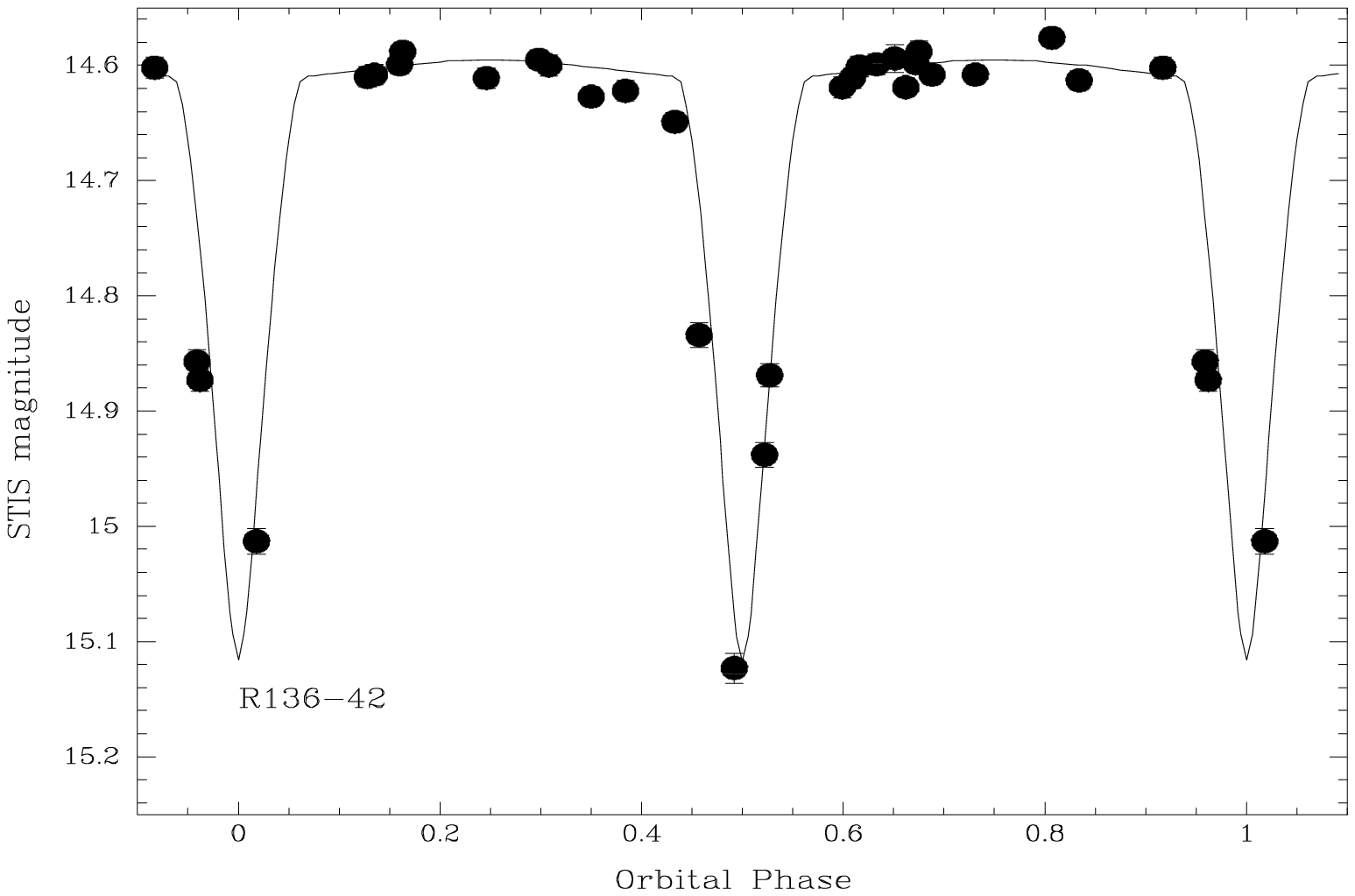}

\caption{Radial velocity and light curves for R136-42; symbols are the
same as in Fig.~\ref{fig:r13638}.  \label{fig:r13642}}
\end{figure}

\clearpage
\begin{figure}
\epsscale{0.7}
\plotone{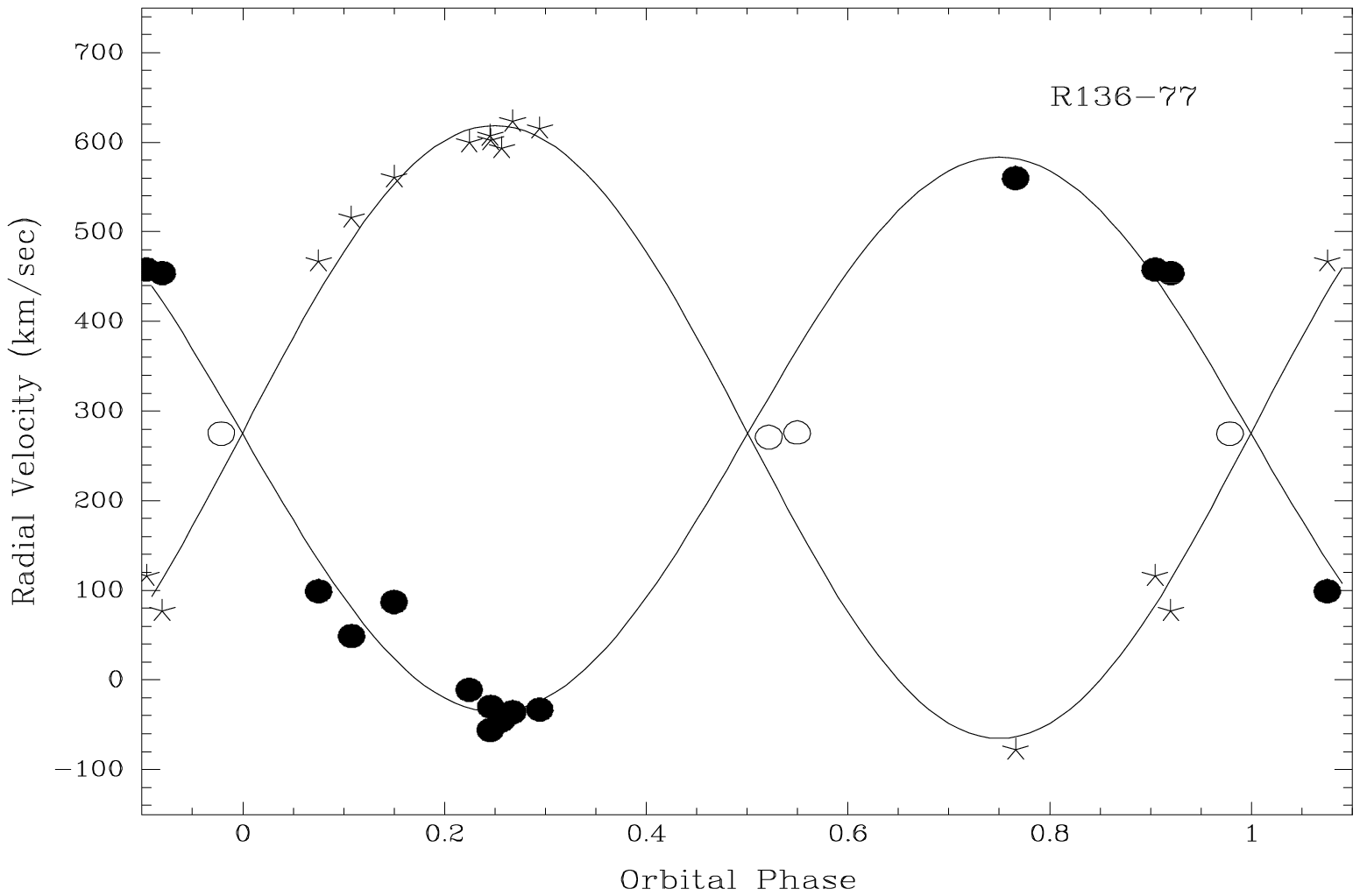}
\plotone{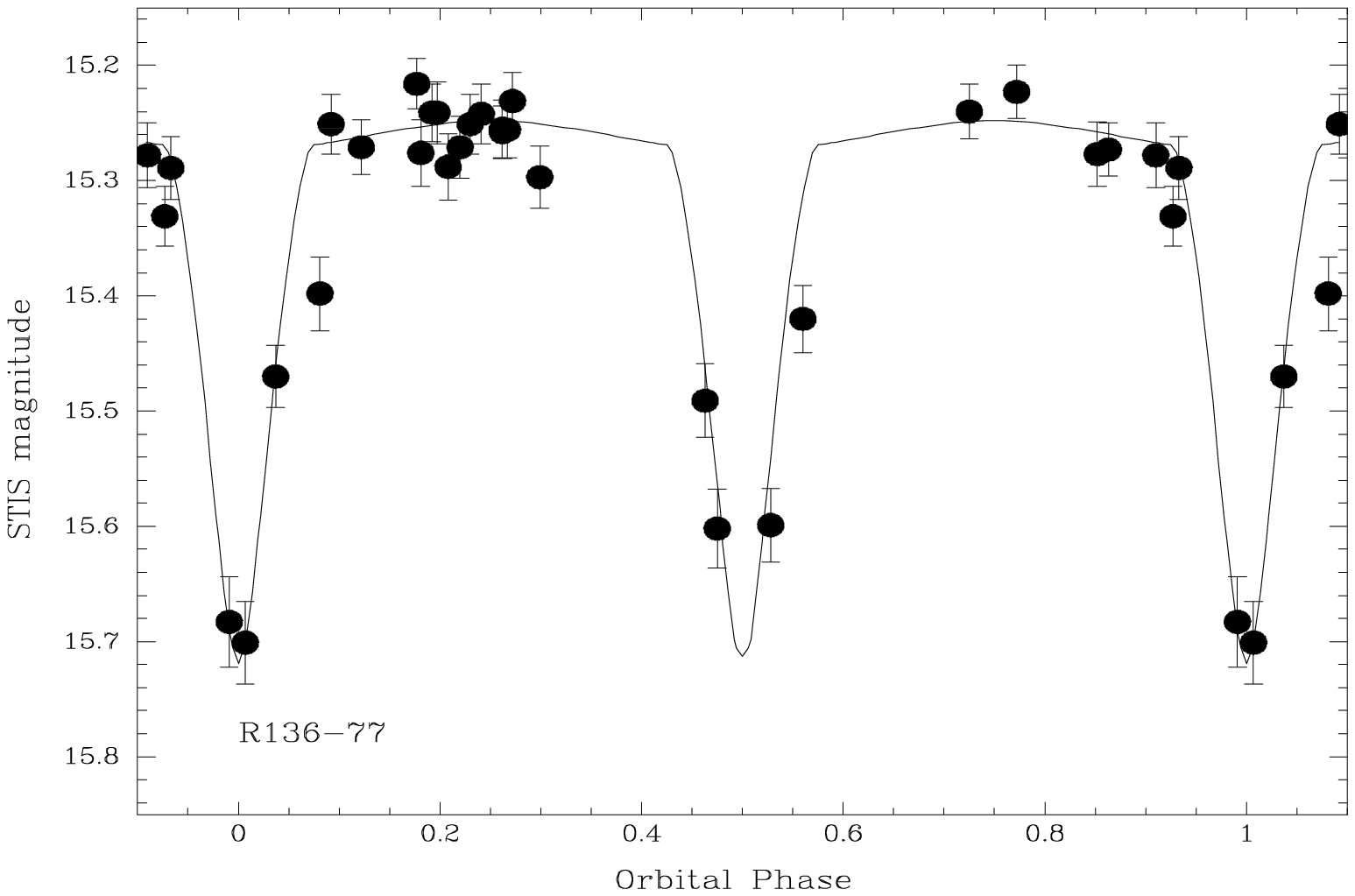}

\caption{Radial velocity and light curves for R136-77; symbols are the
same as in Fig.~\ref{fig:r13638}.  \label{fig:r13677}}
\end{figure}

\clearpage
\begin{figure}
\epsscale{0.48}
\plotone{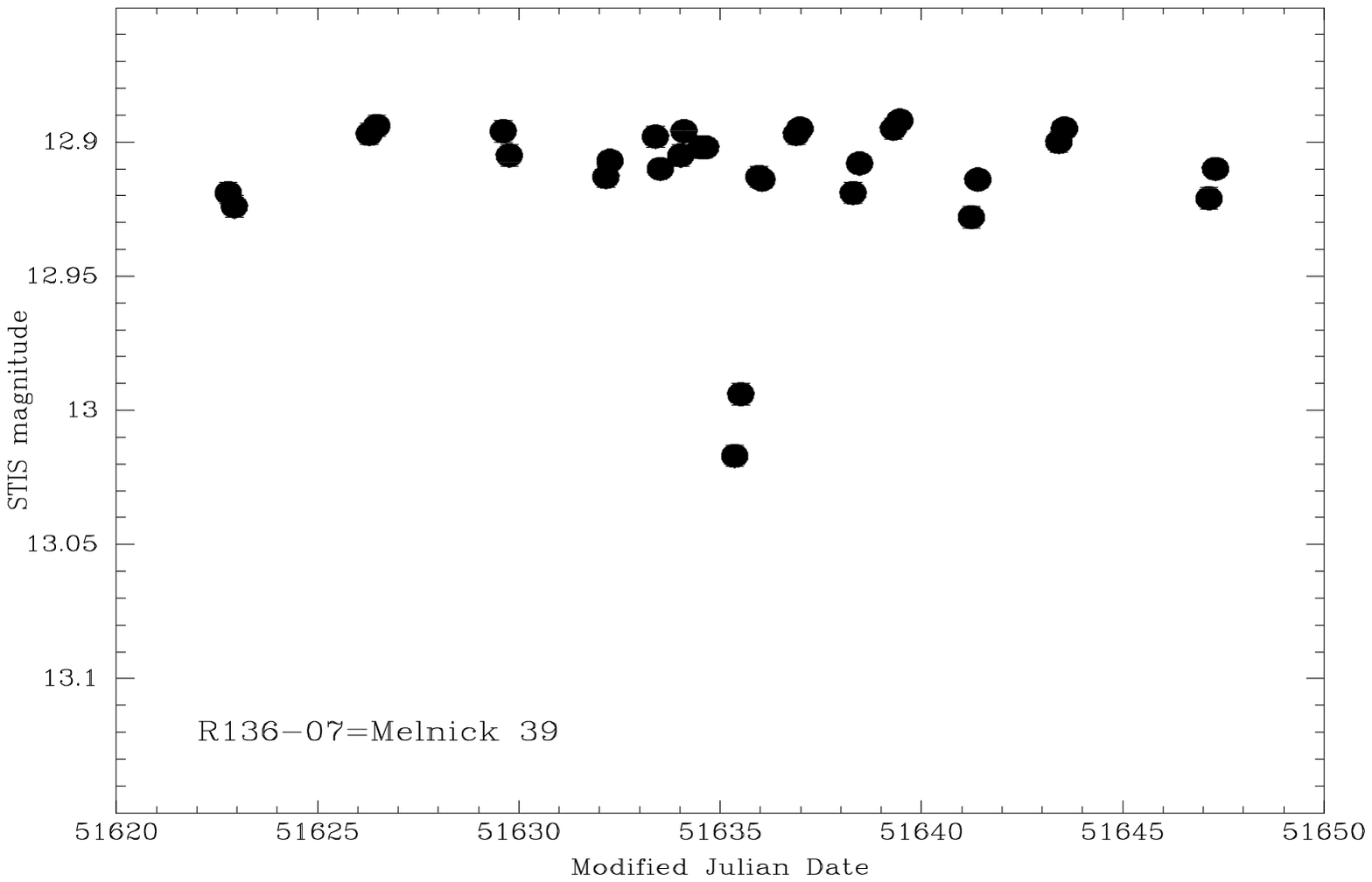}
\plotone{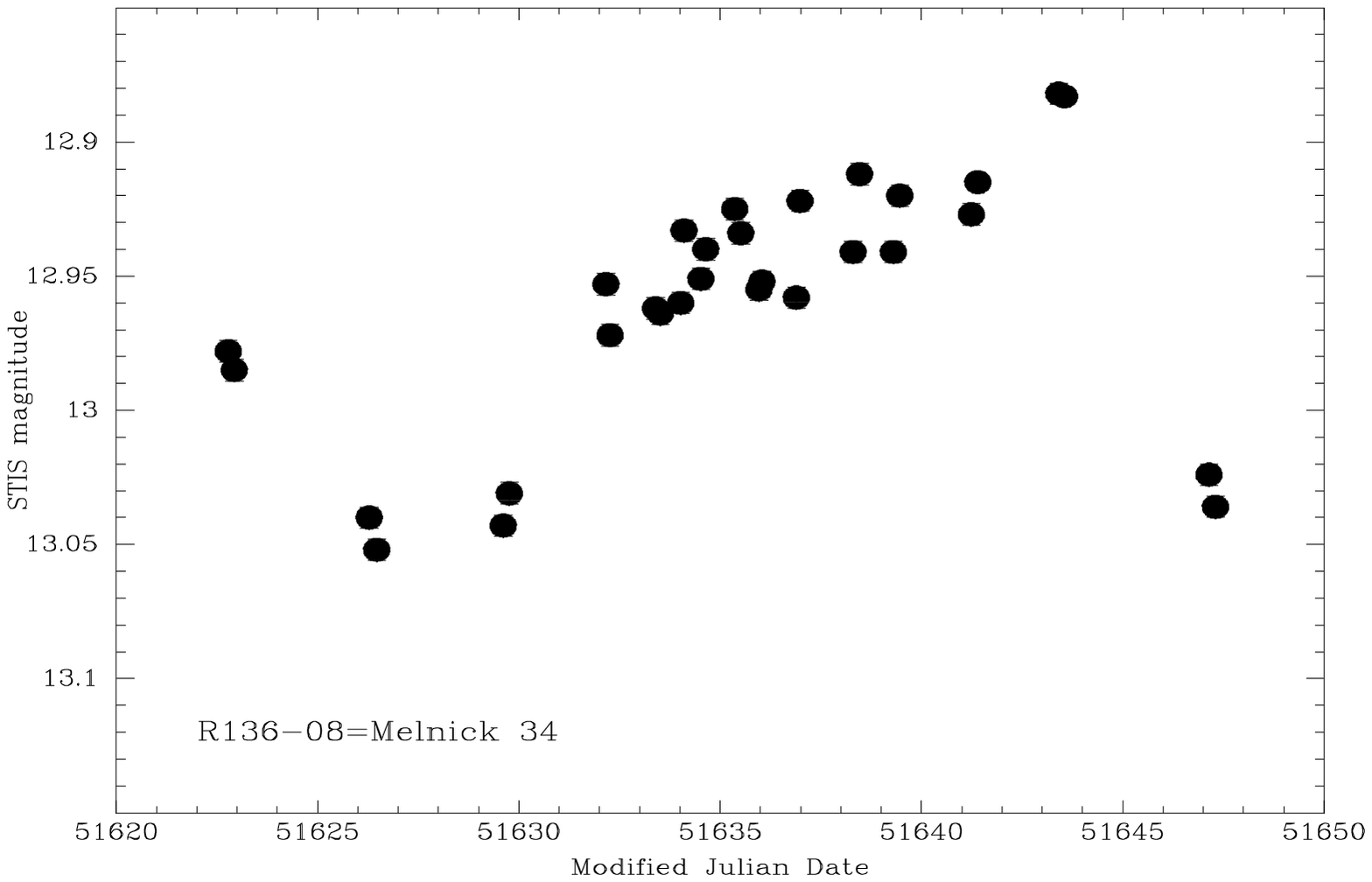}
\plotone{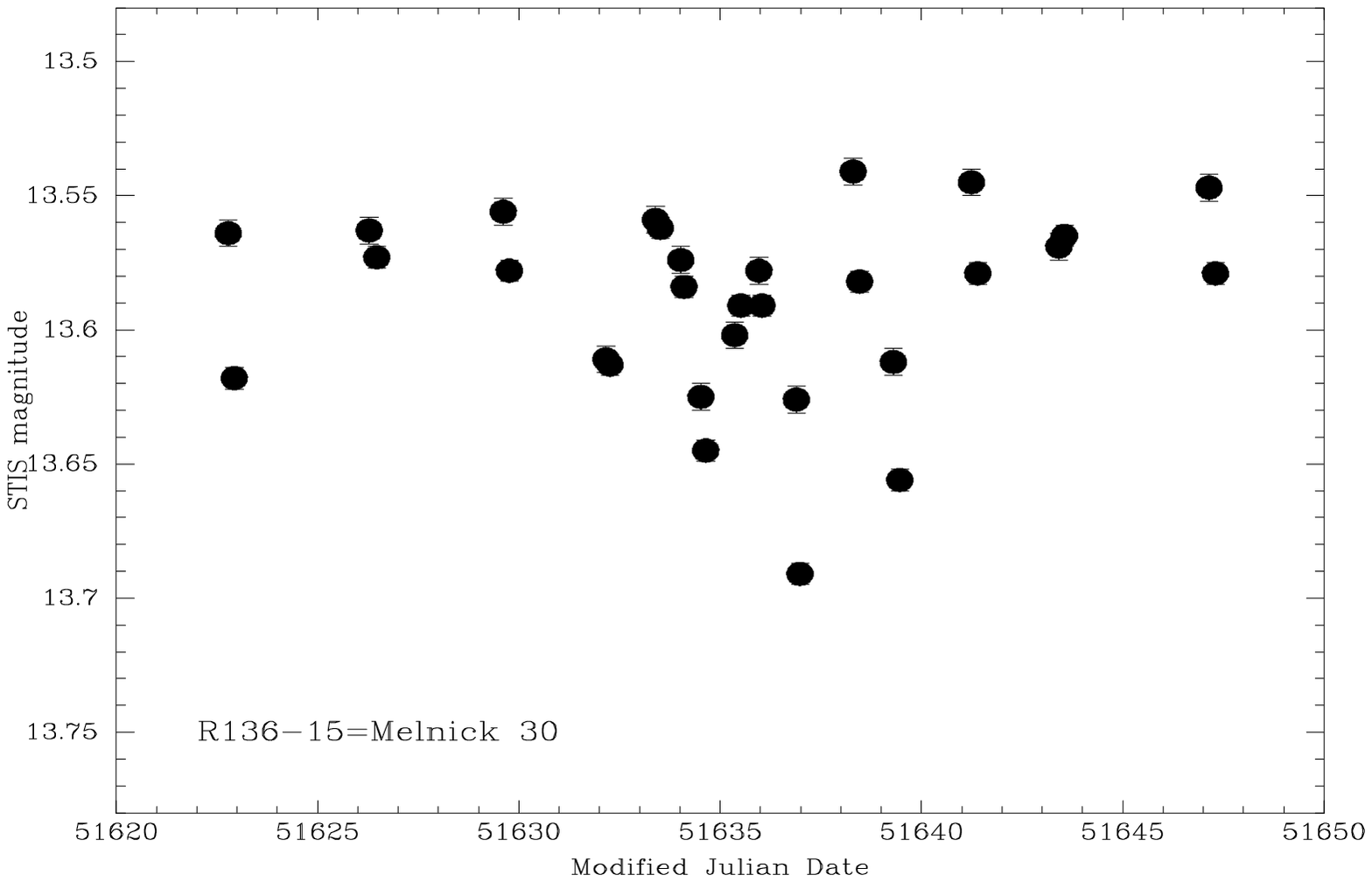}
\plotone{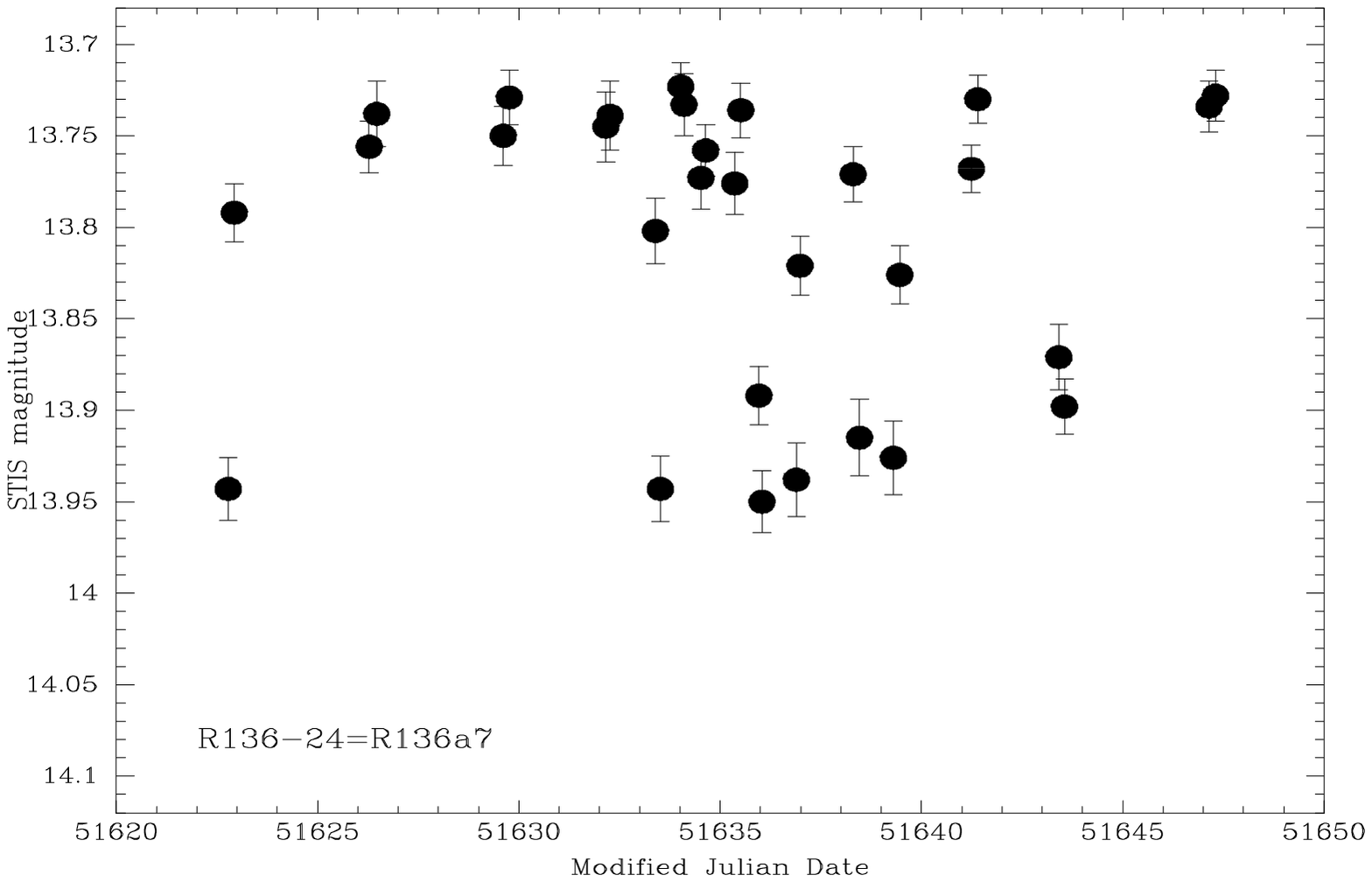}
\plotone{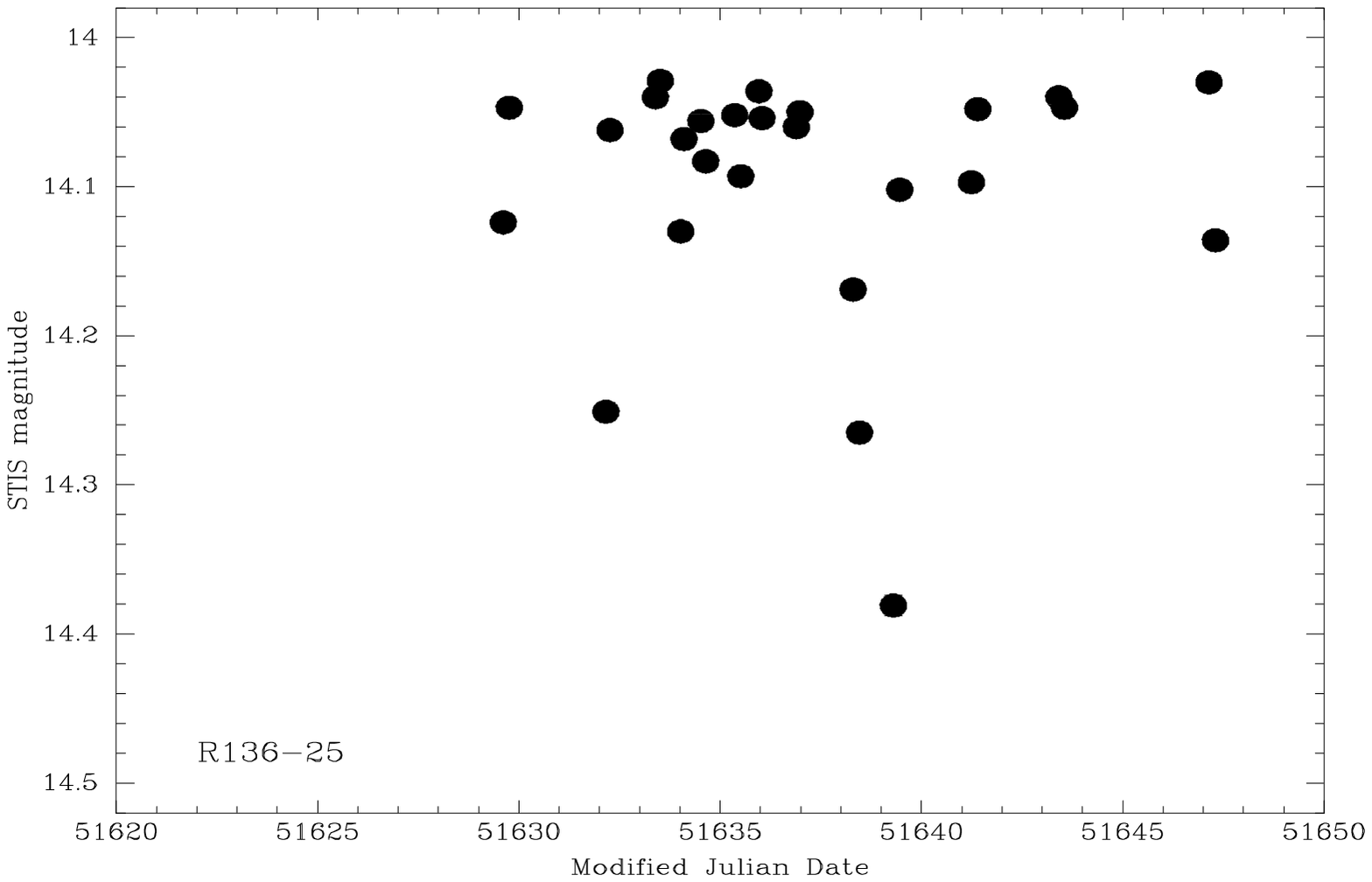}

\caption{Light-curves for five other suspected binaries.  R136-08 (Melnick 34)
show gradual variations with a period of 20 days; the others show 
changes typical of eclipses.  \label{fig:otherlcs}}
\end{figure}

\clearpage
\begin{figure}
\epsscale{0.48}
\plotone{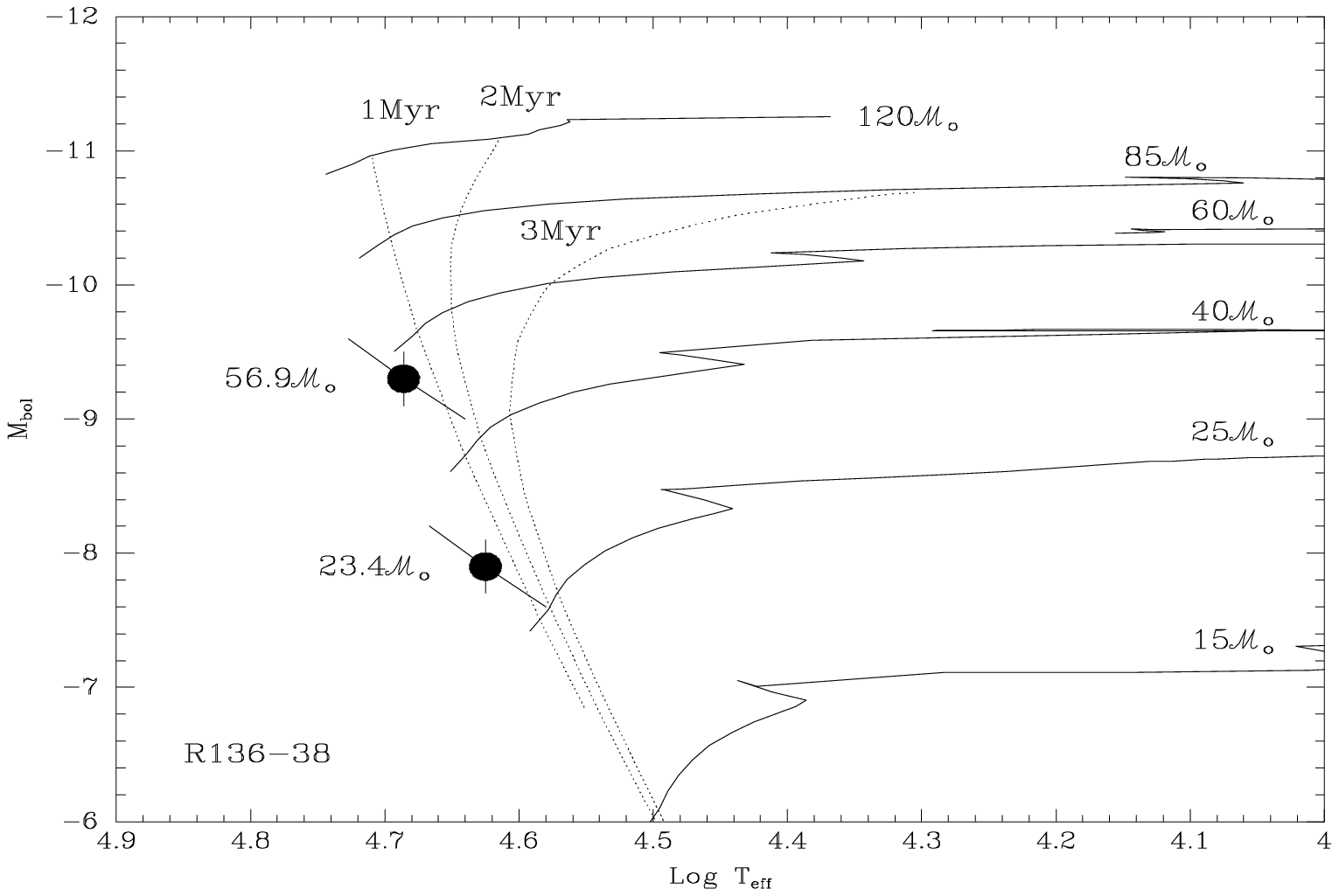}
\plotone{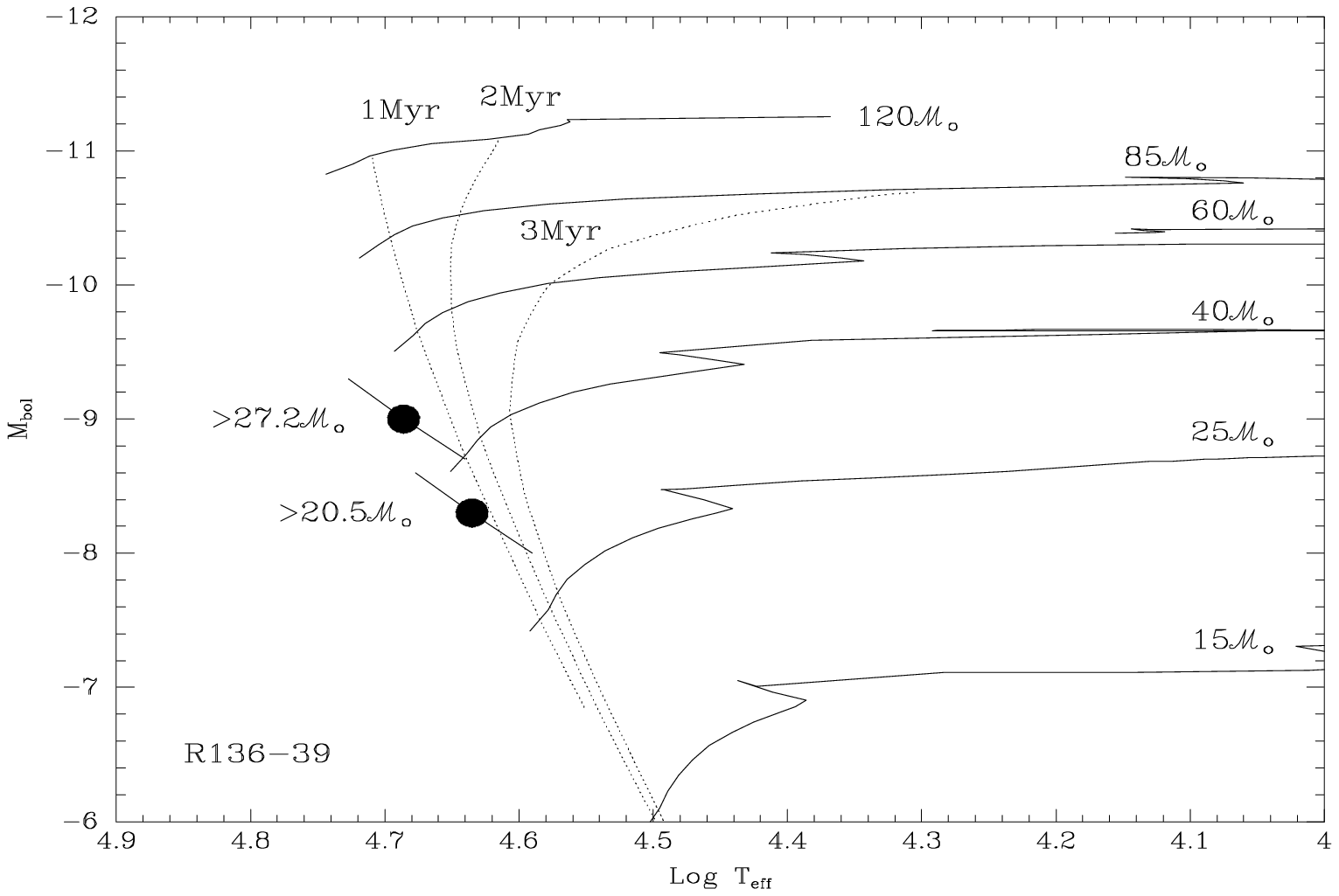}
\plotone{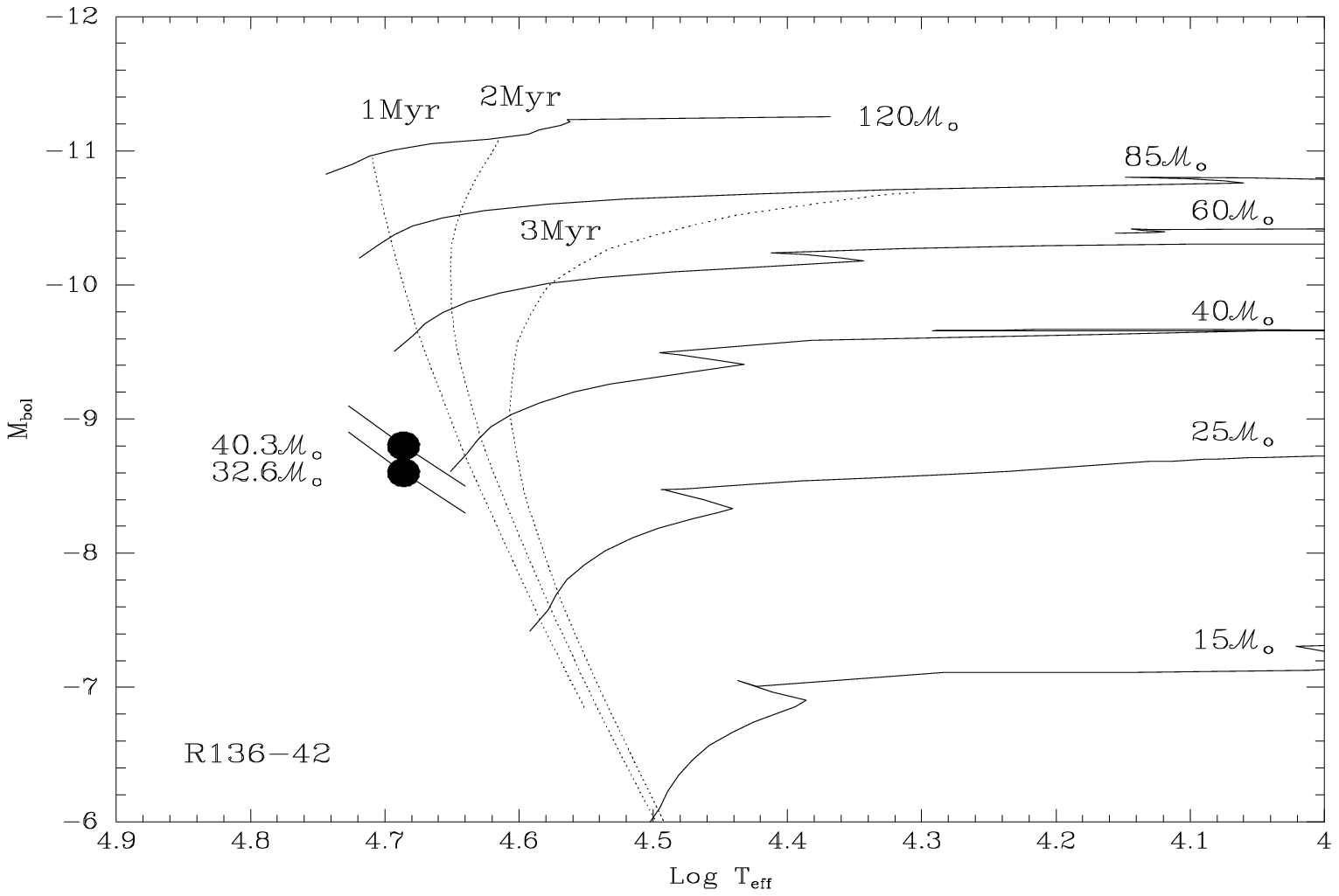}
\plotone{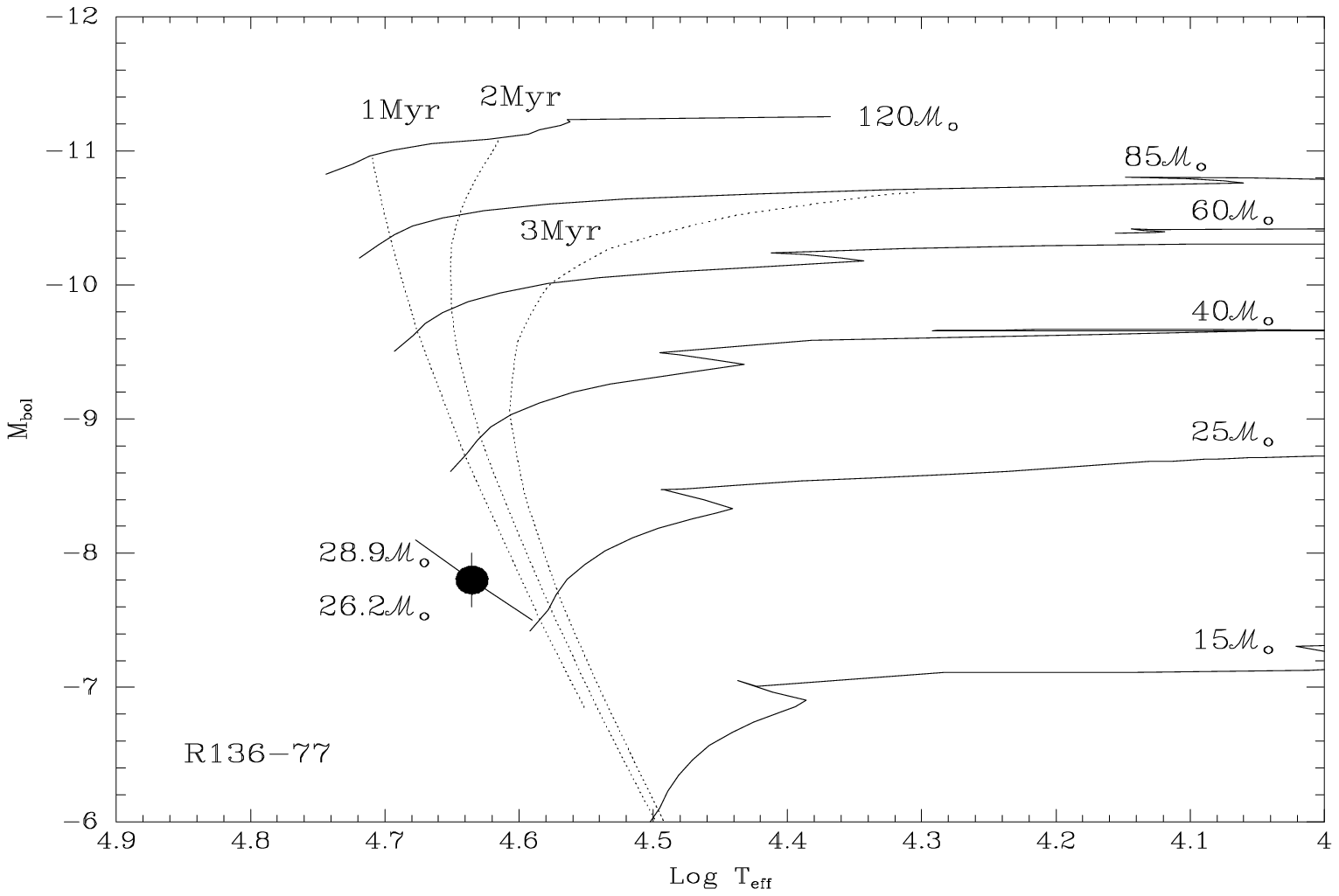}

\caption{The locations of each star in the H-R diagram is shown by
by the large filled circles.  Error bars denote the change in location due to a 5\% error in the effective temperature scale, and due to uncertainties in magnitude difference in the two components.  The evolutionary tracks (solid
lines) are for $z=0.008$ and come from Schaerer et al.\ (1993).  The dotted
lines are isochrones for 1, 2, and 3~Myr, computed from the same models. \label{fig:r136allhrd}}
\end{figure}

\begin{deluxetable}{l c c r r r } 
\tablewidth{0pc}
\tablenum{1}
\tablecolumns{6}
\tablecaption{Journal of Observations, Radial Velocities}
\tablehead{
\colhead{Image}
&\colhead{Modified}
&\colhead{Orbital}
&\multicolumn{3}{c}{Radial Velocities (km s$^{-1}$)} \\ \cline{4-6}
\colhead{Name}
&\colhead{Julian Date}
&\colhead{Phase\tablenotemark{b}}
&\colhead{Primary}
&\colhead{Secondary}
&\colhead{Single}
}
\startdata
\cutinhead{R136-38}
o59001020 & 51622.80 & 0.41 &\nodata & 548 & 202 \\
o59002020 & 51626.29 & 0.44 &\nodata & 473 & 219 \\
o59003020 & 51629.63 & 0.43 &\nodata & 486 & 219 \\
o59004020 & 51632.17 & 0.18 &118  & 626 & \nodata \\ 
o59005020 & 51633.40 & 0.54 &\nodata & \nodata & 300 \\
o59006020 & 51634.03 & 0.73 &450 & $-$118\tablenotemark{a} & \nodata \\
o59007020 & 51634.55 & 0.88 &\nodata & $-$32 & 403 \\
o59008020 & 51635.39 & 0.13 &145 & 581 & \nodata  \\
o59009020 & 51635.97 & 0.30 &112 & 650 & \nodata \\
o59010020 & 51636.91 & 0.58 &\nodata & \nodata & 308 \\
o59011020 & 51638.31 & 0.99 &\nodata & 242 & 279 \\
o59012020 & 51639.32 & 0.29 &117 & 647 & \nodata \\
o59013020 & 51641.25 & 0.86 &418 & $-$96 & \nodata \\
o59014020 & 51643.45 & 0.51 &\nodata & \nodata & 265 \\
o59015020 & 51647.15 & 0.60 &\nodata & 19 & 332 \\
\cutinhead{R136-39}
o59001030 & 51622.84 & 0.56&\nodata & 190 & 262 \\
o59002030 & 51626.34 & 0.42& 141\tablenotemark{a} & 412 & \nodata \\
o59003030 & 51629.67 & 0.24&80 & 514 & \nodata \\
o59004030 & 51632.20 & 0.86&429 & 53 & \nodata \\
o59005030 & 51633.42 & 0.16&102 & 481 & \nodata \\
o59006030 & 51634.04 & 0.31&97 & 509 & \nodata \\
o59007030 & 51634.56 & 0.44&\nodata & 343 & 218 \\
o59008030 & 51635.42 & 0.65&429 & 57 & \nodata \\
o59009030 & 51635.98 & 0.79&454 & 9 & \nodata \\
o59010030 & 51636.92 & 0.02&\nodata & 288\tablenotemark{a} & 251 \\
o59011030 & 51638.36 & 0.38&120 & 446 & \nodata \\
o59012030 & 51639.36 & 0.62&431 & 69 & \nodata \\
o59013030 & 51641.31 & 0.10&148 & 448 & \nodata \\
o59014030 & 51643.46 & 0.63&425 & 45 & \nodata \\
o59015030 & 51647.21 & 0.56&355\tablenotemark{a} & 148\tablenotemark{a} \\
\cutinhead{R136-42}
o59001040 & 51622.86 & 0.27 & 2 & 578 & \nodata \\
o59002040 & 51626.34 & 0.48 &\nodata & \nodata & 250 \\
o59003040 & 51629.69 & 0.64 &468 & 15 & \nodata \\
o59004040 & 51632.20 & 0.51 &\nodata & \nodata & 279 \\
o59005040 & 51633.45 & 0.94 &\nodata & \nodata & 290 \\
o59006040 & 51634.05 & 0.15 &57 & 548 & \nodata \\
o59007040 & 51634.53 & 0.31 &18 & 586 & \nodata \\
o59008040 & 51635.43 & 0.62 &470 & $-$7 & \nodata \\
o59009040 & 51635.99 & 0.82 &529 & $-$29 & \nodata \\
o59010040 & 51636.93 & 0.14 &56 & 535 & \nodata \\
o59011040 & 51638.38 & 0.64 &505 & $-$2 & \nodata \\
o59012040 & 51639.38 & 0.99 &\nodata & \nodata & 274 \\
o59013040 & 51641.39 & 0.69&530 & $-$32 & \nodata \\
o59014040 & 51643.48 & 0.41&128 & 496 & \nodata \\
o59015040 & 51647.22 & 0.70&552 & $-$73 & \nodata \\
\cutinhead{R136-77}
o59001050 & 51622.92 & 0.29& $-$33 & 615 & \nodata \\
o59002050 & 51626.41 & 0.15&   87 & 561 & \nodata \\
o59003050 & 51629.74 & 0.92&  454 &  77 & \nodata \\
o59004050 & 51632.23 & 0.25&$-$30 & 602 & \nodata \\
o59005050 & 51633.47 & 0.90&  458 & 116 & \nodata \\
o59006050 & 51634.07 & 0.22&$-$11 & 600 & \nodata \\
o59007050 & 51634.63 & 0.52& \nodata & \nodata & 271 \\
o59008050 & 51635.49 & 0.98& \nodata & \nodata & 275 \\
o59009050 & 51636.03 & 0.27& $-$36 & 623 & \nodata \\
o59010050 & 51636.97 & 0.77& 560 & $-$78 & \nodata \\
o59011050 & 51638.44 & 0.55& \nodata & \nodata & 276 \\
o59012050 & 51639.43 & 0.08& 99 & 467 & \nodata  \\
o59013050 & 51641.37 & 0.11&49 & 516 & \nodata \\
o59014050 & 51643.53 & 0.26&$-$45 & 593 & \nodata \\
o59015050 & 51647.27 & 0.25&$-$56 & 607 & \nodata \\
\enddata
\tablenotetext{a}{Given half weight in orbit solution.}
\tablenotetext{b}{Based upon the orbital parameters given subsequently.}
\end{deluxetable}

\begin{deluxetable}{l c c r r r } 
\tablewidth{0pc}
\tablenum{2}
\tablecolumns{6}
\tablecaption{Journal of Observations, Photometry}
\tablehead{
\colhead{Image Name}
&\colhead{MJD}
&\colhead{R136-38\tablenotemark{a}}
&\colhead{R136-39\tablenotemark{b}}
&\colhead{R136-42\tablenotemark{c}}
&\colhead{R136-77\tablenotemark{d}}
}
\startdata
o59001010& 51622.78& 14.39& 14.61& 14.61& 15.27\\
o59001060& 51622.93& 14.36& 14.61& 14.60& 15.30\\
o59002010 & 51626.28& 14.35& 14.62& 14.83& 15.40\\
o59002060 & 51626.47& 14.51& 14.60& 14.94& 15.28\\
o59003010 & 51629.61& 14.34& 14.60& 14.61& 15.28\\
o59003060 & 51629.76& 14.39& 14.59& 14.62& 15.29\\
o59004010 & 51636.89& 14.38& 14.63& 14.61& 15.24\\
o59004060 & 51636.98& 14.34& 14.61& 14.60& 15.22\\
o59005010 & 51638.30& 14.54& 14.61& 14.60& 15.60\\
o59005060 & 51638.46& 14.47& 14.60& 14.60& 15.42\\
o59006010 & 51639.30& 14.32& 14.59& 14.87& 15.70\\
o59006060 & 51639.46& 14.32& 14.60& 15.01& 15.25\\
o59007010 & 51641.24& 14.35& 14.59& 14.60& 15.47\\
o59007060 & 51641.40& 14.34& 14.62& 14.61& 15.27\\
o59008010 & 51643.41& 14.47& 14.61& 14.62& 15.24\\
o59008060 & 51643.55& 14.49& 14.60& 14.65& 15.26\\
o59009010& 51647.14& 14.36& 14.60& 14.59& 15.22\\
o59009060& 51647.30& 14.32& 14.60& 14.61& 15.26\\
o59010010& 51632.16& 14.35& 14.61& 15.12& 15.29\\
o59010060& 51632.26& 14.33& 14.60& 14.87& 15.26\\
o59011010& 51633.39& 14.49& 14.61& 14.60& 15.27\\
o59011060& 51633.51& 14.36& 14.60& 14.86& 15.33\\
o59012010& 51634.02& 14.34& 14.60& 14.61& 15.24\\
o59012060& 51634.10& 14.32& 14.62& 14.59& 15.24\\
o59013010& 51634.52& 14.33& 14.60& 14.60& 15.49\\
o59013060& 51634.64& 14.34& 14.61& 14.63& 15.60\\
o59014010& 51635.36& 14.34& 14.61& 14.62& 15.28\\
o59014060& 51635.51& 14.33& 14.61& 14.59& 15.68\\
o59015010& 51635.96& 14.35& 14.59& 14.58& 15.25\\
o59015060& 51636.04& 14.33& 14.61& 14.61& 15.23\\
\enddata
\tablenotetext{a}{The photometric error for R366-38 is 1$\sigma = 0.007$ mag.}
\tablenotetext{b}{The photometric error for R366-39 is 1$\sigma = 0.009$ mag.}
\tablenotetext{c}{The photometric error for R366-42 is 1$\sigma = 0.010$ mag.}
\tablenotetext{c}{The photometric error for R366-77 is 1$\sigma = 0.027$ mag.}
\end{deluxetable}

\begin{deluxetable}{l c c c}
\tablewidth{0pc}
\tablenum{3}
\tablecolumns{4}
\tablecaption{Orbit Solutions and Physical Parameters R136-38}
\tablehead{
\colhead{Parameter}
&\colhead{System}
&\colhead{Primary}
&\colhead{Secondary}
}
\startdata
\sidehead{Orbital}
$P$ (days) & 3.39 (adopted)& \nodata & \nodata\\
$e$        & 0.00 (adopted)& \nodata & \nodata\\
$T_{\rm p.~conjunction}$ (MJD) & $51621.40\pm0.03$  & \nodata & \nodata\\
$\gamma$ (km s$^{-1}$)&\nodata &      $278.2\pm0.4$ &     $272.3\pm0.3$\\
$K$ (km s$^{-1}$)    &\nodata &      $174.7\pm0.5$ &      $424.8\pm0.2$\\
$m_p/m_s$ & 0.41 & \nodata & \nodata \\
R1 (km s$^{-1}$)      &\nodata & 2.3     &   20.3 \\
$a \sin i$ ($R_o$)  &   40.2  & 11.7 & 28.5  \\
$m \sin^3 i$ ($\cal M_\odot$) &\nodata&   $53.8\pm0.2$   & $22.1\pm0.1$ \\
\sidehead{Spectral and Photometric}
Spectral Type       &\nodata          &O3~V    &        O6~V\\
$T_{\rm eff}$($^\circ$K)&\nodata     & $48,500\pm4,850$\tablenotemark{a}    &     $42,200\pm4,220$\tablenotemark{a} \\
$\Delta m $ & $1.0\pm 0.2$ & \nodata & \nodata \\
$M_{V}$  &    $-5.3$   & $-4.9\pm0.2$\tablenotemark{b} & $-3.9\pm0.2$\tablenotemark{b} \\             
BC  &\nodata& $-4.4\pm0.3$\tablenotemark{a}   & $-4.0\pm0.3$\tablenotemark{a} \\
$M_{bol}$& \nodata &   $-9.3\pm0.4$\tablenotemark{a,b}&$-7.9\pm0.4$\tablenotemark{a,b} \\ 
Radius ($R_o$)& \nodata &    $9.3\pm1.0$\tablenotemark{a,b} & $6.4\pm0.7$ \\ 
$v_{\rm sync}$ & \nodata & $110\pm12$ & $76\pm9$ \\
$v \sin i$  & \nodata & $130\pm20$ & $90\pm 20$ \\
Eclipse depths &\nodata &   0.23  &   0.20 \\ 
inclination (geometry) [deg] & $79.0\pm1.0$  & \nodata & \nodata\\
inclination (GENSYN model) [deg] & $79.0\pm1.0$ & \nodata & \nodata \\
\sidehead{Masses}
m ($\cal M_\odot$) orbit  &\nodata&  $56.9\pm0.6$  &   $23.4\pm0.2$ \\
m ($\cal M_\odot$) evolutionary tracks & \nodata & $53\pm5\tablenotemark{c}$ & 
$29\pm2\tablenotemark{c}$ \\
 
\enddata
\tablenotetext{a}{Adopting a 10\% uncertainity in the spectral-type to
$T_{\rm eff}$ relationship.}
\tablenotetext{b}{Errors (anti-)correlated.}
\tablenotetext{c}{Errors on the masses from the evolutionary tracks are based
solely on the errors in $M_V$.}
\end{deluxetable}

\begin{deluxetable}{l c c c}
\tablewidth{0pc}
\tablenum{4}
\tablecolumns{4}
\tablecaption{Orbit Solutions and Physical Parameters R136-39}
\tablehead{
\colhead{Parameter}
&\colhead{System}
&\colhead{Primary}
&\colhead{Secondary}
}
\startdata
\sidehead{Orbital}
$P$ (days) & 4.06 (adopted)& \nodata & \nodata\\
$e$        & 0.00 (adopted)& \nodata & \nodata\\
$T_{\rm p.~conjunction}$ (MJD) & $51620.59\pm0.01$  & \nodata & \nodata\\
$\gamma$ (km s$^{-1}$)&\nodata &      $271.8\pm0.3$ &     $262.0\pm0.3$\\
$K$ (km s$^{-1}$)    &\nodata &      $200.8\pm0.5$ &      $266.3\pm0.4$\\
$m_p/m_s$ & $0.754\pm0.002$ & \nodata & \nodata \\
R1 (km s$^{-1}$)      &\nodata & 7.1     &   9.5 \\
$a \sin i$ ($R_o$)  &   37.5  & 16.1 & 21.4  \\
$m \sin^3 i$ ($\cal M_\odot$) &\nodata&   $24.5\pm0.1$   & $18.5\pm0.1$ \\
\sidehead{Spectral and Photometric}
Spectral Type       &\nodata          &O3~V    &        O5.5~V\\
$T_{\rm eff}$($^\circ$K)&\nodata     & $48,500\pm4,850$\tablenotemark{a}    &     $43,200\pm4,320$\tablenotemark{a} \\
$\Delta m $ & $0.45\pm 0.1$ & \nodata & \nodata \\
$M_{V}$  &    $-5.2$   & $-4.7\pm0.1$\tablenotemark{b} & $-4.2\pm0.1$\tablenotemark{b} \\                
BC  &\nodata& $-4.4\pm0.3$\tablenotemark{a}   & $-4.1\pm0.3$\tablenotemark{a} \\
$M_{bol}$& \nodata &   $-9.0\pm0.3$\tablenotemark{a,b}&$-8.3\pm0.3$\tablenotemark{a,b} \\
Radius ($R_o$)& \nodata &    $8.1\pm0.6$\tablenotemark{a,b} & $7.1\pm0.6$ \\   
$v_{\rm sync}$ & \nodata & $83\pm6$ & $71\pm6$ \\
$v \sin i$ & \nodata & $<100$ & $<100$ \\ 
Eclipse depths &\nodata &   $<0.05?$ & \nodata \\
inclination (geometry)[deg]  & $<72.0?$  & \nodata & \nodata\\
inclination (GENSYN model) [deg] & $<75.0$ & \nodata & \nodata \\
\sidehead{Masses}
m ($\cal M_\odot$) orbit &\nodata&  $>27.2$  &   $>20.5$ \\
m ($\cal M_\odot$) evolutionary tracks & \nodata &$46\pm2\tablenotemark{c}$ & $34\pm2\tablenotemark{c}$ \\
\enddata
\tablenotetext{a}{Adopting a 10\% uncertainity in the spectral-type to
$T_{\rm eff}$ relationship.}
\tablenotetext{b}{Errors (anti-)correlated.}
\tablenotetext{c}{Errors on the masses from the evolutionary tracks are based
solely on the errors in $M_V$.}
\end{deluxetable}

\begin{deluxetable}{l c c c}
\tablewidth{0pc}
\tablenum{5}
\tablecolumns{4}
\tablecaption{Orbit Solutions and Physical Parameters R136-42}
\tablehead{
\colhead{Parameter}
&\colhead{System}
&\colhead{Primary}
&\colhead{Secondary}
}
\startdata
\sidehead{Orbital}
$P$ (days) & 2.89 (adopted)& \nodata & \nodata\\
$e$        & 0.00 (adopted)& \nodata & \nodata\\
$T_{\rm p.~conjunction}$ (MJD) & $51622.07\pm0.01$  & \nodata & \nodata\\
$\gamma$ (km s$^{-1}$)&\nodata &      $276.0\pm0.3$ &     $268.0\pm0.3$\\
$K$ (km s$^{-1}$)    &\nodata &       $278.2\pm0.4$ &      $343.7\pm0.4$\\
$m_p/m_s$ & $0.809\pm0.001$ & \nodata & \nodata \\
R1 (km s$^{-1}$)      &\nodata & 4.7     &   12.8 \\
$a \sin i$ ($R_o$)  &   35.5  & 15.9 & 19.6  \\
$m \sin^3 i$ ($\cal M_\odot$) &\nodata&   $39.9\pm0.1$   & $32.3\pm0.1$ \\
\sidehead{Spectral and Photometric}
Spectral Type       &\nodata          &O3~V    &        O3~V\\
$T_{\rm eff}$($^\circ$K)&\nodata     & $48,500\pm4,850$\tablenotemark{a}    &     $48,500\pm4,850$\tablenotemark{a} \\
$\Delta m $ & $0.2\pm 0.1$ & \nodata & \nodata \\
$M_{V}$  &    $-5.1$   & $-4.4\pm0.1$\tablenotemark{b} & $-4.2\pm0.1$\tablenotemark{b} \\                
BC  &\nodata& $-4.4\pm0.3$\tablenotemark{a}   & $-4.4\pm0.3$\tablenotemark{a} \\
$M_{bol}$& \nodata &   $-8.8\pm0.3$\tablenotemark{a,b}&$-8.6\pm0.3$\tablenotemark{a,b} \\
Radius ($R_o$)& \nodata &    $7.4\pm0.8$\tablenotemark{a,b} & $6.7\pm0.7$ \\   
$v_{\rm sync}$ & \nodata & $102\pm9$ & $93\pm7$ \\
$v \sin i$ & \nodata & $100\pm 20$ & $130\pm 30$ \\
Eclipse depths &\nodata &   $>0.40$ & 0.52 \\
inclination (geometry) [deg] & $85.5\pm1.0$  & \nodata & \nodata\\
inclination (GENSYN model) [deg] &  $85.4\pm0.5$ &\nodata & \nodata \\
\sidehead{Masses}
m ($\cal M_\odot$) orbit &\nodata&  $40.3\pm0.1$  &   $32.6\pm0.1$ \\
m ($\cal M_\odot$) evolutionary tracks & \nodata&  $42\pm2\tablenotemark{c}$ & $39\pm3\tablenotemark{c}$\\
\enddata
\tablenotetext{a}{Adopting a 10\% uncertainity in the spectral-type to
$T_{\rm eff}$ relationship.}
\tablenotetext{b}{Errors (anti-)correlated.}
\tablenotetext{c}{Errors on the masses from the evolutionary tracks are based
solely on the errors in $M_V$.}
\end{deluxetable}

\begin{deluxetable}{l c c c}
\tablewidth{0pc}
\tablenum{6}
\tablecolumns{4}
\tablecaption{Orbit Solutions and Physical Parameters R136-77}
\tablehead{
\colhead{Parameter}
&\colhead{System}
&\colhead{Primary}
&\colhead{Secondary}
}
\startdata
\sidehead{Orbital}
$P$ (days) & 1.88 (adopted)& \nodata & \nodata\\
$e$        & 0.00 (adopted)& \nodata & \nodata\\
$T_{\rm p.~conjunction}$ (MJD) & $51622.37\pm0.01$  & \nodata & \nodata\\
$\gamma$ (km s$^{-1}$)&\nodata &      $273.8\pm0.3$ &     $276.6\pm0.3$\\
$K$ (km s$^{-1}$)    &\nodata &       $309.2\pm0.4$ &      $341.9\pm0.3$\\
$m_p/m_s$ & $0.904\pm0.001$ & \nodata & \nodata \\
R1 (km s$^{-1}$)      &\nodata & 17.5     &   14.8 \\
$a \sin i$ ($R_o$)  &   24.1  & 11.5 & 12.7  \\
$m \sin^3 i$ ($\cal M_\odot$) &\nodata&   $28.3\pm0.1$   & $25.6\pm0.1$ \\
\sidehead{Spectral and Photometric}
Spectral Type       &\nodata          &O5.5~V    &        O5.5~V\\
$T_{\rm eff}$($^\circ$K)&\nodata     & $43,200\pm4,320$\tablenotemark{a}    &     $43,200\pm4,320$\tablenotemark{a} \\
$\Delta m $ & $0.0\pm 0.2$ & \nodata & \nodata \\
$M_{V}$  &    $-4.5$   & $-3.7\pm0.2$\tablenotemark{b} & $-3.7\pm0.2$\tablenotemark{b} \\                
BC  &\nodata& $-4.1\pm0.3$\tablenotemark{a}   & $-4.1\pm0.3$\tablenotemark{a} \\
$M_{bol}$& \nodata &   $-7.8\pm0.4$\tablenotemark{a,b}&$-7.8\pm0.4$\tablenotemark{a,b} \\
Radius ($R_o$)& \nodata &    $5.8\pm0.5$\tablenotemark{a,b} & $5.8\pm0.5$ \\   
$v_{\rm sync}$ & \nodata & $124\pm13$ & $124\pm13$ \\
$v \sin i$ & \nodata &$140\pm20$ & $130\pm20$ \\  
Eclipse depths &\nodata &   $0.45$ & $>0.35$ \\
inclination (geometry) [deg] & $83.0\pm1.0$  & \nodata & \nodata\\
inclination (GENSYN model) [deg]  &  $83.0\pm1.0$  & \nodata & \nodata \\
\sidehead{Masses}
m ($\cal M_\odot$) orbit &\nodata&  $28.9\pm0.3$  &   $26.2\pm0.3$ \\
m ($\cal M_\odot$) evolutionary tracks & \nodata & $28\pm1\tablenotemark{c}$ & $28\pm0.1\tablenotemark{c}$ \\
\enddata
\tablenotetext{a}{Adopting a 10\% uncertainity in the spectral-type to
$T_{\rm eff}$ relationship.}
\tablenotetext{b}{Errors (anti-)correlated.}
\tablenotetext{c}{Errors on the masses from the evolutionary tracks are based
solely on the errors in $M_V$.}
\end{deluxetable}


\clearpage
\begin{references}

\reference {} Al-Naimiy, H. M. 1978, Ap\&SS, 53, 181

\reference {} 
Antokhina, E. A., Moffat, A. F. J., Antokhin, I. I., Bertrand, J.-F.,
\& Lamontagne, R. 2000, ApJ, 529, 463

\reference {} Bagnuolo, W. G., Gies, D. R., \& Wiggs, M. S. 1992, ApJ, 385,
708.


\reference {} Bevington, P. R. 1969, Data Reduction and Error Analysis
for the Physical Sciences (New York: McGraw-Hill)

\reference {} Brunish, W. M., \& Truran, J. W. 1982, ApJ, 256, 247

\reference {} Burkholder, V., Massey, P., \& Morrell, N. 1997, ApJ, 490, 328

\reference {} Chlebowski, T., \& Garmany, C. D. 1991, ApJ, 368, 241

\reference {} Claret, A. 2000, A\&A, 363, 1081

\reference {} Conti, P. S. 1988, in O Stars and Wolf-Rayet Stars., ed.~P. S.
Conti \& A. B. Underhill, NASA SP-497 (Washington, DC: NASA), 124

\reference {} Heger, A., \& Langer, N. 2000, ApJ, 544, 1016

\reference {} Heger, A., Langer, N., \& Woosley, S. E. 2000, ApJ, 528, 368

\reference {} Herrero, A., Kudritzki, R. P., Vilchez, J. M., Kunze, D., Butler,
K., \& Haser, S.\ 1992, A\&A, 261, 209

\reference {} Herrero, A., Puls, J., \& Villamariz, M. R. 2000, A\&A, 354, 193

\reference {} Hunter, D. A., Vacca, W. D., Massey, P., Lynds, R., \& O'Neil, E. J. 1997, AJ, 113, 1691

\reference {} Kudritzki, R.-P., \& Puls, J. 2000, ARA\&A, 38, 613

\reference {} Kurucz, R. L. 1979, ApJS, 40, 1

\reference {} Lafler, J., \& Kinman, T. D. 1965, ApJS, 11, 216

\reference {} Leitherer et al.\ 2001, STIS Instrument Handbook, Version 5.1 (Baltimore: STScI)

\reference {} Massey, P. 1998, in 
The Stellar Initial Mass Function, 38th Herstmonceux Conference, 
ed.\ G. Gilmore \& D. Howell, (San Francisco, ASP), 17 

\reference {} Massey, P., \& Conti, P. S. 1977, ApJ, 218, 431

\reference {} Massey, P., \& Hunter, D. A. 1998, ApJ, 493, 180

\reference {} Massey, P., DeGioia-Eastwood, K., and Waterhouse, E. 2001,  
AJ, 121, 1050 

\reference {} Massey, P., Waterhouse, E., and DeGioia-Eastwood, K. 2000, AJ,
119, 2214 

\reference {} Maeder, A., \& Conti, P. S. 1994, ARA\&A, 32, 227

\reference {} Maeder, A., \& Meynet, G. 2000, ARA\&A, 38, 143

\reference {} Maeder, A., \& Meynet, G. 2001, A\&A, 373, 555

\reference {} Meynet, G., \& Maeder, A. 2000, A\&A, 361, 101

\reference {} Mochnacki, S. W., \& Doughty, N. A. 1972, MNRAS, 156, 51

\reference {} Penny, L. R., Gies, D. R., \& Bagnuolo, W. G. 1999, in
Wolf-Rayet Phenomena in Massive Stars and Starburst Galaxies, ed.
K. A. van der Hucht, G. Hoenigsberger, \& P. R. J. Eenens (San Francisco, ASP),
86

\reference {} Puls, J., Kudritzki, R. P., Herrero, A., Pauldrach, A. W. A.,
Haser, S. M., Lennon, D. J., Gabler, R., Voels, S. A., Vilchez, J. M., 
Wachter, S., \& Feldmeier, A. 1996, A\&A, 305, 171

\reference {} Schaerer, D., Meynet, G., Maeder, A., \&Schaller, G. 1993,
A\&AS, 98, 523

\reference {} Schaller, G., Schaerer, D., Meynet, G., \& Maeder, A. 1992,
A\&AS, 96, 269

\reference {} Schweickhardt, J., Schmutz, W., Stahl, O., Szeigfert, T.,
\& Wolf, B. 1999, A\&A, 347, 127

\reference {} Shu, F. H., \& Lubow, S. H. 1981, ARA\&A 19, 277

\reference {} Vacca, W. D., Garmany, C. D., \& Shull, J. M. 1996, ApJ, 460, 914

\reference {} van den Bergh, S. 2000, The Galaxies of the Local Group
(Cambridge: Cambridge University Press)

\reference {} van Hamme, W. 1993, AJ, 106, 2096

\reference {} Westerlund, B. E. 1997, The Magellanic Clouds (Cambridge:
Cambridge University Press)

\reference {} Wolfe, R. H., Horak, H. G., \& Storer, N. W. 1967, in
Modern Astrophysics, ed.\ M. Hack (Paris: Gauthier Villars), 251

\reference {} Zahn, J.-P. 1975, A\&A, 41, 329

\reference {} Zahn, J.-P. 1977, A\&A, 57, 383

\end{references}
\end{document}